\documentclass[graybox]{svmult}

\usepackage{type1cm}        
\usepackage{makeidx}         
\usepackage{graphicx}        
\usepackage{multicol}        
\usepackage[bottom]{footmisc}

\usepackage{newtxtext}       %
\usepackage{newtxmath}       

\newcommand{\BF}[1]{{\color{black} #1}}
\usepackage{amsmath} 

\makeindex             

\begin{document}

\title*{Noncovalent Interactions of Hydrated DNA and RNA Mapped by 2D-IR Spectroscopy}
\author{Benjamin P. Fingerhut and Thomas Elsaesser}
\institute{B. P. Fingerhut \at Max-Born-Institute, 2a Max-Born-St., Berlin, 12489, Germany \email{fingerhut@mbi-berlin.de}
\and T. Elsaesser \at Max-Born-Institute, 2a Max-Born-St., Berlin, 12489, Germany \email{elsasser@mbi-berlin.de}}
%
%
\maketitle

\abstract{Biomolecules couple to their aqueous environment through a variety of noncovalent interactions. Local structures at the surface of DNA and RNA are frequently determined by hydrogen bonds with water molecules, complemented by non-specific electrostatic and many-body interactions. Structural fluctuations of the water shell result in fluctuating Coulomb forces on polar and/or ionic groups of the biomolecular structure and in a breaking and reformation of hydrogen bonds. Two-dimensional infrared (2D-IR) spectroscopy of vibrational modes of DNA and RNA gives insight into local hydration geometries, elementary molecular dynamics, and the mechanisms behind them. In this chapter, recent results from 2D-IR spectroscopy of native and artificial DNA and RNA are presented, together with theoretical calculations of molecular couplings and molecular dynamics simulations. Backbone vibrations of DNA and RNA are established as sensitive noninvasive probes of the complex behavior of hydrated helices. The results reveal the femtosecond fluctuation dynamics of the water shell, the short-range character of Coulomb interactions, and the strength and fluctuation amplitudes of interfacial electric fields.
}

\abstract*{Each chapter should be preceded by an abstract (no more than 200 words) that summarizes the content. The abstract will appear \textit{online} at \url{www.SpringerLink.com} and be available with unrestricted access. This allows unregistered users to read the abstract as a teaser for the complete chapter.\newline\indent
Please use the 'starred' version of the \texttt{abstract} command for typesetting the text of the online abstracts (cf. source file of this chapter template \texttt{abstract}) and include them with the source files of your manuscript. Use the plain \texttt{abstract} command if the abstract is also to appear in the printed version of the book.}

\section{Vibrational Probes of Interactions and Dynamics in Aqueous Systems}
\label{sec:1}
Liquid water represents the medium in which most biological processes occur. The structure of bulk water at ambient temperature, its structural fluctuations, and their impact on chemical and biological processes  have been the subject of extensive experimental and theoretical research over the last decades. On the experimental side, vibrational spectroscopy with femtosecond time resolution has played a key role in identifying and separating the different contributions to water dynamics in the electronic ground state of the bulk liquid. In most experiments, excitations of the OH stretching and, to lesser extent, OH bending vibration of the water molecule have served for probing structure fluctuations, intermolecular energy exchange, as well as hydrogen bond breaking and reformation. Two-dimensional infrared (2D-IR) spectroscopy has allowed for extracting frequency fluctuation correlation functions (FFCFs) of the liquid, in which the different types of molecular dynamics give rise to different decay components. Moreover, vibrational relaxation processes and energy dissipation in the liquid have been studied both for intra- and intermolecular water modes, including the transient librational response. Theoretical calculations and molecular dynamics simulations have addressed the relevant molecular interactions and driving forces behind the different processes. In the present context, we briefly summarize a few key aspects of the dynamics of bulk H$_2$O, more detailed information can be found in recent review articles
\cite{Bakker:ChemRev:2010,Nibbering:ChemRev:2004,Laage:ChemRev:2017}.

Water molecules in bulk H$_2$O form, on the average, four hydrogen bonds (H bonds), two in which the OH groups act as hydrogen donors, and two in which the oxygen atom serves as hydrogen acceptor. At ambient temperature, molecular degrees of freedom in a wide frequency range below the intramolecular OH bending and stretching modes are thermally excited, leading to stochastic motions of water molecules on a multitude of time scales. Such structural fluctuations give rise to fluctuating electric fields which originate from the dipolar character of the water molecule with a liquid-phase electric dipole moment of approximately 2.9 Debye \cite{Silvestrelli:PRL:1999}. The fastest fluctuations occur in the sub-100 fs time domain and are connected with librational motions essentially localized on individual water molecules. Slower subpicosecond structure changes are due to delocalized librational degrees of freedom as well as hydrogen bond stretching and bending motions. The lifetime of hydrogen bonds is on the order of 1 ps, resulting in a rapid change of the extended hydrogen bond network in the liquid. According to the so-called jump model, the breaking and reformation of H bonds is connected with a jump-like angular reorientation of water molecules induced by fluctuations in the fourfold coordination of water molecules \cite{Laage:Science:2006}. The rotational reorientation time of water molecules has a value of approximately 2.5 ps. All such processes result in a decay of structural correlations between water molecules.  

The intramolecular OH stretching and bending modes of H$_2$O display lifetimes of some 200 fs. For OH stretching excitations, intermolecular excitation transfer between neighboring water molecules has been observed on a time scale of 100 fs. The dissipation of vibrational excess energy involves a transfer from the intramolecular vibrations to librations of a sub-100 fs lifetime and a subsequent redistribution among low-frequency degrees of freedom. As a result, a quasi-thermal ground state characterized by a vibrational excess temperature is formed within 1 to 2 ps \cite{Ashihara:JPCA:2007}. 

A biomolecule embedded in water introduces both steric boundary conditions and specific interaction sites for water molecules in the first and second layer of the hydration shell. Vice versa, the degree of hydration has a marked impact on the structure of biomolecules, e.g., on the particular double-helical form of DNA \cite{Saenger:Nature:1986}. 
Recent years have witnessed a change in perception of these first few water layers, from a passive bystander to an active player in determining structural stability and dynamics of biological entities.
Equilibrium geometries of DNA and RNA have extensively been studied by x-ray diffraction and positions of water oxygens and counterions, typically for crystallized samples at a limited hydration level, have been identified
\cite{Drew:JMB:1981,Vlieghe:AC:1999,Schneider:BJ:1998,Egli:BP:1998}. In DNA and RNA structures, the ionic phosphate groups of the sugar-phosphate backbone represent primary hydration sites at which each of the two free oxygen atoms accepts up to three hydrogen bonds with first-layer water molecules. The sugar OH groups and NH and carbonyl groups of nucleic bases can form hydrogen bonds to water as well. While the first water layer displays a substantial structural heterogeneity, the impact of the biomolecule on the hydration shell structure is limited to a few, typically less than 5 water layers \cite{Laage:ChemRev:2017}. It should be noted that x-ray diffraction preferentially maps immobilized water molecules at particular interaction sites while mobile waters have remained elusive.

Noncovalent interactions and water dynamics in hydration shells are currently understood only in part. In particular, the dynamics of hydrating water have remained controversial. A dramatic slowing down of structural fluctuations and rotational motions due to a more rigid water structure has been claimed on the basis of solvation studies in which electronic chromophores were attached to or incorporated in biomolecular structures \cite{Pal:ChemRev:2004,Zhong:CPL:2011,Andreatta:2005aa}. After electronic excitation by an ultrashort optical pulse, the transient red-shift of emission has been followed to derive the FFCF of the aqueous environment. Such FFCFs display kinetic components extending into the nanosecond range and beyond which have been assigned to water dynamics. In contrast, nuclear magnetic resonance, ultrafast infrared studies, and molecular dynamics simulations point to a moderate slowing down of water dynamics in hydration shells by a factor of 3 to 5 only, with the fastest components well in the femtosecond time domain \cite{Halle:JPCB:2009,Furse:2008aa,Yang:JPCB:2011}. Obviously, the different experimental methods monitor different aspects of the highly complex dynamic scenario. 
Moreover, some of the probes are invasive, i.e., they impact solvation geometries and molecular arrangements substantially.

Insight into the spatial extent, dynamic nature and microscopic origins of the coupling of water with the biomolecule provides a basis for a deeper understanding of biomolecular interactions which are governed by the hydration / de-hydration forces at the interface.
Functional  groups of biomolecules which are located at the interface to the water shell hold particular potential for probing interactions, in particular hydrogen bonding and electrostatic couplings, as well as  hydration dynamics. In DNA and RNA, vibrations of the helix backbone represent noninvasive probes which interact with the water shell and embedded counterions \cite{Siebert:JPCB:2015}. Two-dimensional infrared (2D-IR) spectroscopy \cite{HammZanni,ChoBook} with a femtosecond time resolution gives insight into the coupled dynamics of biomolecules and their water shells, into anharmonic couplings and excitation transfer between backbone vibrations, and into energy exchange between the two subsystems \cite{Siebert:JPCB:2015,Guchhait:StructDyn:2016,Siebert:JPCL:2016,Liu:StructDyn:2017}. This method together with in-depth theory and simulation has successfully been applied to native and artificial DNA and RNA structures and to model systems providing specific information on particular types of interactions \cite{Costard:JCP:2015,Costard:PCCP:2015,Fingerhut:JCP:2016}. In this chapter, we discuss recent results from such research. After a brief introduction in the methods (section 2), 2D-IR spectra of artificial DNA and RNA oligomers at different levels of hydration are presented in section 3, including an analysis of 2D lineshapes and the highly complex intermode coupling patterns. Section 4 discusses electric fields at the DNA surface, their fluctuations and their spatial range. Conclusions are given in section 5, together with an outlook on how to map interactions of DNA and RNA with ions in their environment.   

\begin{figure}
\centering
\includegraphics[scale=1.2]{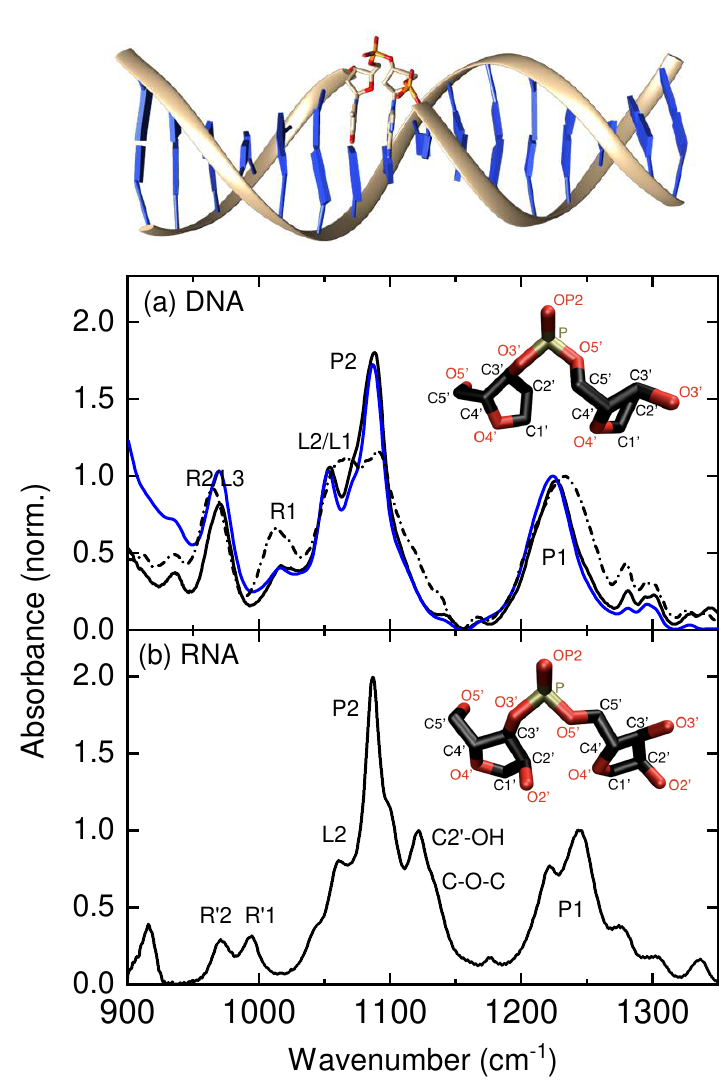}
%
%
\caption{Top panel: Schematic of a B-DNA double helix with two backbone strands (grey) and base pairs (blue). (a) Infrared absorption bands of DNA backbone modes. The absorbance A=-log(T) (T: sample transmission) is plotted as a function of wavenumber for double-stranded oligomers containing 23 alternating adenine-thymine pairs at a hydration level of 92\% relative humidity (r.h., black dash-dotted line) and for full hydration (black solid line). Blue line: absorption spectrum of fully hydrated salmon testes DNA with some 2000 base pairs. The absorbance is normalized to the peak of the asymmetric PO$_2^-$ stretching band P1. The other bands are due to the symmetric PO$_2^-$ vibration P2, the ribose vibrations R$_{1,2}$ and the linker modes L$_{1,2,3}$. Inset: backbone segment of B-DNA. (b) Backbone absorption spectrum of a fully hydrated RNA double stranded helix containing 23 adenine-uracil base pairs. Inset: backbone segment of an A-RNA helix. }
\label{fig:1}       
\end{figure}      

\section{Experimental and Theoretical Methods}
\label{sec:2}

This section gives a short description of sample preparation (sec.~\ref{sec:2.1}) and the methods of nonlinear vibrational spectroscopy applied in the present study (sec.~\ref{sec:2.2}). We then describe a hierarchy of theoretical methods that provide an understanding of  the noncolvalent interactions of biomolecules with their surrounding (sec.~\ref{sec:2.3}). A summary on the applicability for various system sizes and time scales is given.

\subsection{Preparation and Linear Infrared Spectra of Hydrated DNA and RNA Samples}
\label{sec:2.1}

Different DNA and RNA structures with different counterions were studied at hydration levels from 0 \% relative humidity (r.h.) corresponding to two water molecules per base pair up to full hydration with more than 150 water molecules per base pair. Here, we focus on results for double stranded DNA oligomers containing 23 alternating adenine-thymine (A-T) base pairs and native salmon testes DNA which contains approximately 2000 base pairs, among them 41 \% guanine-cytosine (G-C) and 59 \% A-T pairs. As a prototypical RNA structure, we use double-stranded oligomers with 23 alternating adenine-uracil (A-U) base pairs in Watson-Crick geometry. Fully hydrated samples were prepared by dissolving the DNA and RNA helices and their Na$^+$ counterions in water or an 0.1 M aqueous NaCl solution. The concentration of salmon testes DNA was on the order of 10$^{-4}$ M while the concentration of the DNA and RNA oligomers was typically 10$^{-2}$M. A sample of a 5 to 10 $\mu$m thickness was cast between two BaF$_2$ or Si$_3$N$_4$ windows. 

The experiments at reduced hydration level were performed with thin-film DNA and RNA samples, prepared by exchanging the Na$^+$ counterions against cetyltrimethylammonium (CTMA), applying the procedure reported in ref.~\cite{Tanaka:JACS:1996}. The DNA and RNA concentration in the films was 10$^{-2}$ M. Films of 10 to 25 $\mu$m thickness were prepared on BaF$_2$ or Si$_3$N$_4$ substrates and placed in a humidity cell \cite{Dwyer:RSI:2013}. This cell was connected to a reservoir containing a saturated aqueous NaBrO$_3$ solution (92 \% r.h., 20 to 30 water molecules per base pair) or P$_5$O$_5$ powder (0\% r.h.). The hydration level was monitored via the spectral position of the asymmmetric PO$_2^-$ stretching vibration of the DNA and RNA backbones \cite{Falk:JACS:1963}. 

Linear infrared spectra of DNA and RNA samples are summarized in Fig. 1. At high hydration levels, salmon testes DNA and the double-stranded DNA oligomers exist in a B-helix structure (top panel of Fig. 1) while double-stranded RNA forms an A-helix.  In the frequency range from 900 to 1350 cm$^{-1}$, there are infrared absorption bands of seven characteristic backbone normal modes of B-DNA (Fig. 1a) \cite{Siebert:JPCB:2015,Banyay:BiophysChem:2003,Guan:Biopol:1996}. The asymmetric (P1) and the symmetric (P2) PO$_2^-$ stretching band of the phosphate groups in the backbone is located around 1220 and 1090 cm$^{-1}$, respectively. The spectral position of P1 depends sensibly on the hydration level, displaying a red-shift with increasing water level. This behavior is evident from the absorption spectra of the DNA oligomers at 92 \% r.h. (dash-dotted line) and for full hydration (black solid line). The phosphate stretching absorption bands are complemented by the absorption of the linker modes L1 to L3 and the ribose modes R1 and R2. A detailed analysis in terms of normal modes has been presented in ref.~\cite{Guan:Biopol:1996}. It should be noted that, apart from minor changes of absorption strength and line shapes, the spectra of fully hydrated oligomers and salmon testes DNA display the same absorption pattern.

The additional OH group of the ribose unit in RNA compared to DNA gives rise to changes in the linear infrared spectrum. The spectrum of the fully hydrated RNA A-helices presented in Fig. 1(b) exhibits moderate spectral shifts of the phosphate and linker modes and a different pattern of the ribose modes R1 and R2 \cite{Bruening:JPCL:2018}. The P1 band of RNA is split into two subcomponents reflecting different local hydration patterns. The additional bands at 1120 and 1135 cm$^{-1}$ are due to the C$_{2'}$-OH stretching mode of the additional OH group and to the C$_{1'}$-O$_{4'}$-C$_{4'}$ ribose stretching mode. The different RNA bands display a somewhat reduced linewidth compared to the corresponding bands in the DNA spectra, pointing to a reduced structural disorder of the hydrated A-helix.

\subsection{Two-Dimensional Infrared Spectroscopy and Pump-Probe Methods}
\label{sec:2.2}

Two-dimensional infrared (2D-IR) and pump-probe spectroscopy are used for mapping the third-order nonlinear response of backbone vibrations of hydrated DNA, RNA, and model systems such as dimethyl phosphate. Heterodyne detected 3-pulse photon echoes are recorded in the standard folded-box-CARS geometry \cite{Cowan:Nature:2005}. In this scheme, 3 femtosecond infrared pulses with wavevectors {\bf k}$_{1,2,3}$ interact sequentially with the sample. The first two pulses are separated by the coherence time $\tau$ while the third pulse follows after the population time T. The four-wave-mixing signal emitted in the direction {\bf k}$_s$=-{\bf k}$_1$ + {\bf k}$_2$ + {\bf k}$_3$ is heterodyned with a fourth local oscillator pulse traveling collinearly with the signal. The resulting interferogram is spectrally dispersed in a monochromator and detected by a 64-element array detector with a 2-cm$^{-1}$ spectral resolution, thus giving the nonlinear signal as a function of detection frequency $\nu_3$. Measurements of the signal as a function of coherence time and a subsequent Fourier transform along $\tau$ provide the nonlinear vibrational response as a function of the excitation frequency $\nu_1$. In this scheme, two pairs of phase-locked pulses are generated with the help of diffractive optics, the pulses 1 and 2 as well as pulse 3 and the local oscillator. Throughout this chapter, we present absorptive 2D signals, given as the real part of the sum of the rephasing and non-rephasing third-order signals \cite{Khalil:JPCA:2003}.

Femtosecond pulses tunable in wide spectral range from 900 to 4000 cm$^{-1}$ were generated by two- or three-stage parametric frequency converters driven by sub-50 fs pulses from amplified Ti:sapphire lasers (repetition rate 1 kHz). The millijoule input pulses are first converted into signal and idler pulses tunable in the near-infrared between 4000 and 8000 cm$^{-1}$, using BBO crystals. In a second step, difference frequency mixing of the near-infrared pulses in GaSe crystals of sub-millimeter thickness provides mid-infrared pulses of 80 to 150 fs duration (depending on the spectral position) and pulses energies up to 7 $\mu$J. Details of the pulse generation and characterization have been given in ref.~\cite{Siebert:JPCB:2015}. All 2D-IR measurements were performed with pulses of parallel linear polarization.    
   
The 2D-IR experiments were complemented by temporally and spectrally resolved pump-probe experiments. Independently tunable pump and probe pulses were generated in two optical parametric amplifiers, the intensity ratio between pump and probe pulses was 50:1 and the time resolution approximately 100 fs \cite{Liu:StructDyn:2017}. Pump and probe pulses were linearly polarized and interact with the sample in a non-collinear transmission geometry. The fraction of excited molecular oscillators was less than 1 \%. After interacting with the sample, both the probe and a reference beam for normalization were spectrally dispersed in a monochromator and detected by a double HgCdTe detector array with 64 elements each. The spectral resolution was 2 cm$^{-1}$. Normalization of the probe intensity to the reference pulse allows for the measurement of pump-induced absorbance changes $\Delta$A down to some $5 \times 10^{-5}$ OD.

\subsection{Theoretical Description of Biomolecular Electrostatics}
\label{sec:2.3}

The modeling of biological systems at the molecular scale requires a description in terms of the atomic interactions. For solvation structures and dynamics around biomolecules the description of electrostatics is of particular importance due to the long-range nature and the pronounced polarity of water  molecules that interact with, e.g., the charged phosphate groups of DNA and RNA. This molecular arrangement defines a highly complex many-body problem which is understood only in part. The  structural dynamics in the first water layers impose electric field fluctuation amplitudes at the phosphate site on a multitude of time scales via, e.g., librational motions and the breaking and reformation of hydrogen bonds. In particular, fluctuation amplitudes and the effective spatial range of the electric fields at the surface of the biomolecule  have remained highly controversial.

At the interface of the biomolecule with its aqueous surrounding  the local static dielectric constant  undergoes changes on molecular length scales.
The arising dielectric discontinuity is an important quantity 
for electrostatic models  rooted in Poisson-Boltzmann theory   \cite{Kirmizialtin_BiophysJ_2012}  and the results depend on the choice of the radial dielectric screening function $\epsilon(r)$~\cite{Young_JPCB_1998,Jayram_Biopol_1989}. For example, the choice of the static end-point of the discontinuity  narrows the molecular contact region and affects the specificity of DNA-counterion interactions \cite{Cuervo_PNAS_2014}.  Moreover, Poisson-Boltzmann models do not account for a molecular description of the solvent and ion-ion correlations are absent. The simulation of solvation structures and dynamics  aiming at the interpretation of multidimensional vibrational spectroscopic signals on a molecular level thus calls for atomistic descriptions. An excellent overview over theoretical methods employed for the description of biomolecular electrostatics can be found in refs.~\cite{Ren:QRevBiophys:2012,Cisneros_ChemRev_2014}.

\subsubsection{Molecular Dynamics Simulations of Solvation Structure and Dynamics}
\label{sec:MD}

Atomistic molecular dynamics simulations mitigate crucial assumptions of Poisson-Boltzmann models and provide a molecular picture of the biomolecular interface via a representation of the molecular electrostatic potential in terms of atomic partial charges. The results crucially depend on quality and reliability of the employed force fields ~\cite{Dans:CurOp:2016}. Observations of accumulating $\alpha/\gamma$ transitions \cite{Bevergidge:BiophysJ:2004} that corrupt DNA duplex structures on the tens-of-nanosecond time scale have spurred substantial effort towards a reparametrization of DNA force field parameters, in particular sugar puckering of the DNA sugar-phosphate backbone (e.g., the BSC0~\cite{Perez:BiophysJ:2007} and BSC1~\cite{Ivani:NatMeth:2015} variants of the AMBER force field and the CHARMM36 force field \cite{Hart:JCTC:2012}). With such improved re-parametrization stability of DNA double strands has been demonstrated on the microsecond time scale \cite{Galindo-Murillo:NatCom:2014} with excellent agreement to experimental 
reference structures \cite{Rodrigo:JCTC:2016,Adam:NucAcidRes:2017}.

While the DNA force fields increasingly became reliable recently, the increased flexibility of RNA poses substantial difficulties and deficiencies have been identified in particular for noncanonical structures \cite{Sponer:JPCL:2014,Sponer:ChemRev:2018}.  Accordingly, force field refinement of RNA is a topic of ongoing research and the reliability of results should be carefully verified by comparison to experimental data.
Similarly, counterion force field parameters are undergoing continuous  reparametrization, in particular for multiple charged ions like  Mg$^{2+}$ and Ca$^{2+}$ \cite{Allner:JCTC:2012,Yoo:JPCL:2012,Martinek:JCP:2018,Li:ChemRev:2017}. Further challenges arise due to large computational requirements on system size, required equilibration times of the ion atmosphere \cite{Lavery:NucAcidRes:2014} and a lack of experimental benchmark data.

Due to recent progress in porting molecular dynamics algorithms to graphical processing unit (GPU) hardware  \cite{Salomon-Ferrer:JCTC:2013,Pall:CPC:2013}  
 atomistic force field simulations  of the electrostatics of solvated biomolecules  today provide routine access to the hundreds of nanoseconds to microsecond time scale for systems like oligomeric  DNA/RNA double strands embedded in explicit water solvent.

\subsubsection{Higher Order Multipoles, Many-Body Polarization Effects and Electrostatic Field-Frequency Mapping}
\label{sec:field}

Pairwise additive force-fields rely on a point-charge representation of the molecular charge density. Such a description  provides access to the bulk polarization of water via a reorientation of individual water molecules subject to an imposed electric field, but is to be distinguished from a rearrangement of electronic structure within the individual molecules  due to many-body polarization effects (induction). Further, higher order charge multipoles are commonly neglected. 
	Major effort has been devoted  to the development of polarizable force fields \cite{Ren_JCPB_2003,Troester_JPCB_2013} with first biomolecular applications approaching maturity \cite{Cisneros_ChemRev_2014,Ponder_JPCB_2010}.  Many-body polarization is expected to be particularly important for the description of water at biological interfaces \cite{Kuo_JPCB_2001,Ploetz_JCTC_2016,Cisneros_ChemRev_2016}.

The Effective Fragment Potential (EFP) model \cite{Day_JCP_1996,Gordon_JCPA_2001,Gordon_ChemRev_2012,Kaliman_JCC_2013} provides a description of intermolecular electrostatics that accounts for many-body polarization and higher order charge multipoles. Within the EFP model  the intermolecular interaction potential $U_{EFP}$ is partitioned into fundamental contributions that each have clear physical origin:
\begin{equation}
U_{EFP}  =  U_{El} + U_{Pol} + U_{Disp} + U_{Xr} + U_{CT}.
\label{eq:EFP}
\end{equation}
Here, $U_{El, Pol, Disp, Xr, CT}$ denote electrostatic, polarization, dispersion, exchange repulsion and charge transfer  contributions, respectively. In the following, we focus primarily on electrostatic and induced dipole interactions (polarization) which are expected to be the dominant contributions  for long range field fluctuations at the charged phosphate-water interface  due to their limiting $1/r$ and $1/r^3$ dependence.

\paragraph{{\bf{Electrostatic Interactions:}}}
The electrostatic intermolecular interaction potential~$U_{El}$ is given by
\begin{align}
U_k^{El} = & q_k \phi(r_x) - \sum_a^{x,y,z} \mu^a_kE_a(r_{kx}) \nonumber \\
& + \frac{1}{3} \sum_{a,b}^{x,y,z} \Theta_k^{ab}E_{ab}(r_{kx})
- \frac{1}{15}  \sum_{a,b,c}^{x,y,z} \Omega_k^{abc}E_{abc}(r_{kx})
\label{eq:Elpot}
\end{align}
where  $\phi(r_x)$  denotes the molecular electrostatic potential expanded up to octopoles\BF{,}
$E_a = \BF{-}\nabla \phi(r_x)$ denotes the electric field, $E_{ab} = \nabla^2 \phi(r_x)$  the electric field gradient and $E_{abc} = \nabla^2E_{a}$ the electric field Hessian and 
$q_k$,  $\mu_k^a$,  $\Theta_k^{ab}$ and $\Omega_k^{abc}$ are  
molecular electric quantities, i.e., charges, dipole, quadrupole and octopole moments, located at expansion points $k$.

\paragraph{{\bf{Polarization Contribution:}}}
 \BF{M}any-body polarization forms a  non-additive part of the total electrostatic interaction that is neglected in treatments with additive force fields. 
 Within the EFP model polarization effects  are considered self-consistently  by mutual induction of dipoles at centroids of localized molecular orbitals 
 due to an imposed electric field ($\mu_{i,\alpha}^{ind}  = \alpha\cdot E$).  The  electric field has components from  static multipoles $M_j$ as well as induced dipoles $\mu_{k,\alpha}^{ind}$ at other sites $k$:
\begin{equation}
\mu_{i,\alpha}^{ind} = \alpha_{i,\alpha,\beta} \left[ \sum_j T_{\alpha}^{ij} M_j + \sum_k T_{\alpha\beta}^{ik} \mu_{k,\beta}^{ind} \right], \quad \alpha,\beta = x,y,z.
\label{eq:inddip}
\end{equation}

\paragraph{{\bf{Electric Field - Frequency Correlation:}}}
Focussing on the electric field $E_a$  as the dominant contribution to the solvatochromic shift of phosphate 
stretch vibrations~\cite{Levinson2011}, the field  is evaluated up to quadrupole interactions:
\begin{align}
E_a(r_x) = \sum_k \left[ q_k T^a(r_{kx}) - \sum_\alpha^{x,y,z} \mu_k^{a,tot} T^{ab}(r_{kx})  \right. \nonumber \\
  \left. + \frac{1}{3} \sum_{b,a}^{x,y,z} \Theta_k^{ab} T^{abc}(r_{kx}) \right].
 \label{eq:fieldEFP}
\end{align}
Here $\mu_k^{a,tot}$ accounts for induced dipoles  and  permanent dipoles ($\mu_k^{a,tot} = \mu_k^{a,ind} + \mu_k^{a,el}$) where the induced dipoles have been converged subject to mutual  polarization (eq.~\ref{eq:inddip}) and $T$ denotes the electrostatic tensors up to rank three. The electric field is evaluated as scalar quantity via a  projection on the symmetry axis of the PO$_2$$^-$ group ($E^{{C2}}$).

Assuming a linear relation, the solvent field induced solvatochromic shift is given by 
\begin{equation}
 \omega=  \omega_0 + \delta\omega = \omega_0  + a\cdot E^{{C2}}
 \label{eq:map}
 \end{equation}
 where $a$  denotes the  slope of the electric field - frequency correlation
 that can be compared to the experimental Stark tuning rate \cite{Fried_AccChemRes_2015,Kim:ChemRev:2013,Basiak:AccChemRes:2017} and $\omega_0$ denotes the field free frequency of, e.g., the asymmetric   PO$_2$$^-$  stretching vibration that is accessible from ab-initio simulations (cf. \BF{s}ec.~\ref{sec:abinitio}).

Parameters of the EFP model can be derived entirely from high level ab-intio simulations. 
The computational scaling is determined by self-consistent determination of induced dipoles (eq.~\ref{eq:inddip}) and allows to access fluctuation dynamics of nanosecond trajectories. For solvated DNA model systems like, e.g., dimethyl phosphate (DMP) converged field fluctuation correlation functions in the $\approx$ 10 ps range, the natural time scale of hydrogen-bond exchange were demonstrated \cite{Fingerhut:JCP:2016}.

\subsubsection{Ab-Initio Simulations of DNA/RNA Backbone Vibrations}
\label{sec:abinitio}

Ab-initio simulations of biomolecular systems interacting with a water environment are prohibitively expensive but provide most accurate insight into noncolvalent intermolecular interactions. Due to the high computational costs, direct molecular dynamics simulations of biomolecules embedded in water are up-to-date not feasible on relevant time scales and carefully chosen model systems are required.
As such ab-initio simulations 
primarily serve for the following purposes:

(1) Benchmark of electric field amplitudes:
the magnitude of EFP simulated electric fields, imposed on the phosphate group by the solvent molecules, can be  rigorously benchmarked in ab-initio simulations of solvated DNA model systems like DMP \cite{Fingerhut:JCP:2016}. Such calculations  (i) provide a benchmark on  the accuracy of  the EFP method and (ii) serve as indicator for dispersion, exchange repulsion and charge transfer  contributions to the total intermolecular interaction, as the  field evaluated with QM methods,  like second order perturbation theory (MP2), naturally contains all contributions to the intermolecular interaction potential (cf. eq.~\ref{eq:EFP}).

(2) Microscopic picture of the molecular vibrational Hamiltonian:
high level, ab-initio derived multidimensional vibrational model Hamiltonians provide insight into details of the molecular Hamiltonian and can be further employed to simulate  linear IR and absorptive 2D-IR spectra of phosphate ion model systems in bulk H$_2$O~\cite{Costard:PCCP:2015}. The ab-initio derived model Hamiltonians reveal that  inter-mode couplings and anharmonicities are generally small (< 15 cm$^{-1}$, see sec.~\ref{sec:3}). Extensive quantum mechanical / molecular mechanical (QM/MM) simulations of backbone vibrations of an RNA (AU)$_{23}$ oligomer were further employed to provide a consistent microscopic mode assignments of the canonical \BF{A-}RNA duplex sugar-phosphate backbone \cite{Bruening:JPCL:2018}. The simulations revealed the delicate impact of the C\BF{$_{2'}$}-OH group on vibrational spectra.  With such information  a microscopic structural picture of mode couplings can be developed. 

(3) Insight into solvatochromic shifts and noncovalent interactions:
DNA model systems like DMP in aqueous solution show pronounced solvent imposed frequency shifts  ($\approx$ 60-80 cm$^{-1}$) of the asymmetric PO$_2^-$ stretching vibration (eq.~\ref{eq:map}) that closely resemble observed shifts of DNA phosphate stretching vibrations at different hydration levels.
Ab-initio normal mode analysis  and vibrational configuration interaction (VCI) simulations  revealed how the phosphate stretch vibrational potential is decisively determined by the noncovalent interactions with the first few water shells \cite{Costard:PCCP:2015,Fingerhut:JCP:2016}. Such simulations provide a rationale how the phosphate vibrational modes serve as sensitive probes of hydration structure in DNA and RNA. 

Ab-initio normal mode analysis 
further revealed that solvent induced frequency shifts of the asymmetric PO$_2^-$ stretching vibration are primarily determined by the electrostatic phosphate-water interactions \cite{Costard:JCP:2015}. Due to the high polarizability, the rearrangement of the electronic structure within the phosphate group is the dominant mechanism via electrostatic interactions while charge transfer contributions are minor. Only upon formation of contact ion pairs (cf. sec~\ref{sec:5}) \cite{Schauss:JPCL:2019}, exchange repulsion interactions become important and impose an $\approx$ 30~cm$^{-1}$ blue shift of the  asymmetric PO$_2^-$ stretching vibration.

\subsubsection{Density Matrix Simulations for the Modeling of Two-Dimensional Infrared Spectra.}

The 2D spectra were analyzed by calculating the third-order response functions to the experimental photon-echo pulse sequence by a perturbative density matrix approach of light-matter interaction, including Kubo lineshape analysis and the lifetime broadening caused by population relaxation of the different vibrations. The treatment follows the perturbative approach developed in ref.~\cite{HammZanni,MukamelBible}, a detailed description of this treatment has been given in refs.~\cite{Siebert:JPCB:2015,Costard:PCCP:2015}. 
The fluctuating forces originating from the aqueous environment and DNA/RNA structure fluctuations are included via the frequency fluctuation correlation function (FFCF). 
The FFCF 
 is given by the Kubo ansatz
\begin{equation}
\langle \delta \nu_i(t) \delta \nu_i(0)\rangle =  \Delta^2_{1,i} \exp (-t/\tau_1) +  \Delta^2_{2,i} \exp (-t/\tau_2)\BF{.}
\label{eq:ffcf}
\end{equation}
 Here, $\delta \nu_i(t)$ denotes the frequency excursion of mode $i$ at time $t$ from its average transition frequency and $\Delta_{1,i}$ and $\Delta_{2,i}$ are fluctuation amplitudes of the frequency excursions, $\tau_1$ and $\tau_2$ are the fluctuation correlation times. 
  To limit the number of free parameters in the modeling the values of $\tau_{1,2}$ are kept fixed in the simulations and the amplitudes $\Delta_{1,i}$ and $\Delta_{2,i}$  are adjusted to best reproduce the  experimental line shapes. 
 In the modeling of 2D spectra (see sec.~\ref{sec:3}) a slow ($\tau_2 \approx$ 50 ps) component is assumed that causes a quasi-static inhomogeneous broadening of the lineshapes and a fast component ($\tau_1 \approx$ 300 fs) accounts for spectral diffusion. Microscopic simulations of the DMP model system provide similar time scales with slightly different values (50 fs, 1.2 ps) \cite{Fingerhut:JCP:2016}.

\begin{figure}[t]
\centering
\includegraphics[scale=0.43]{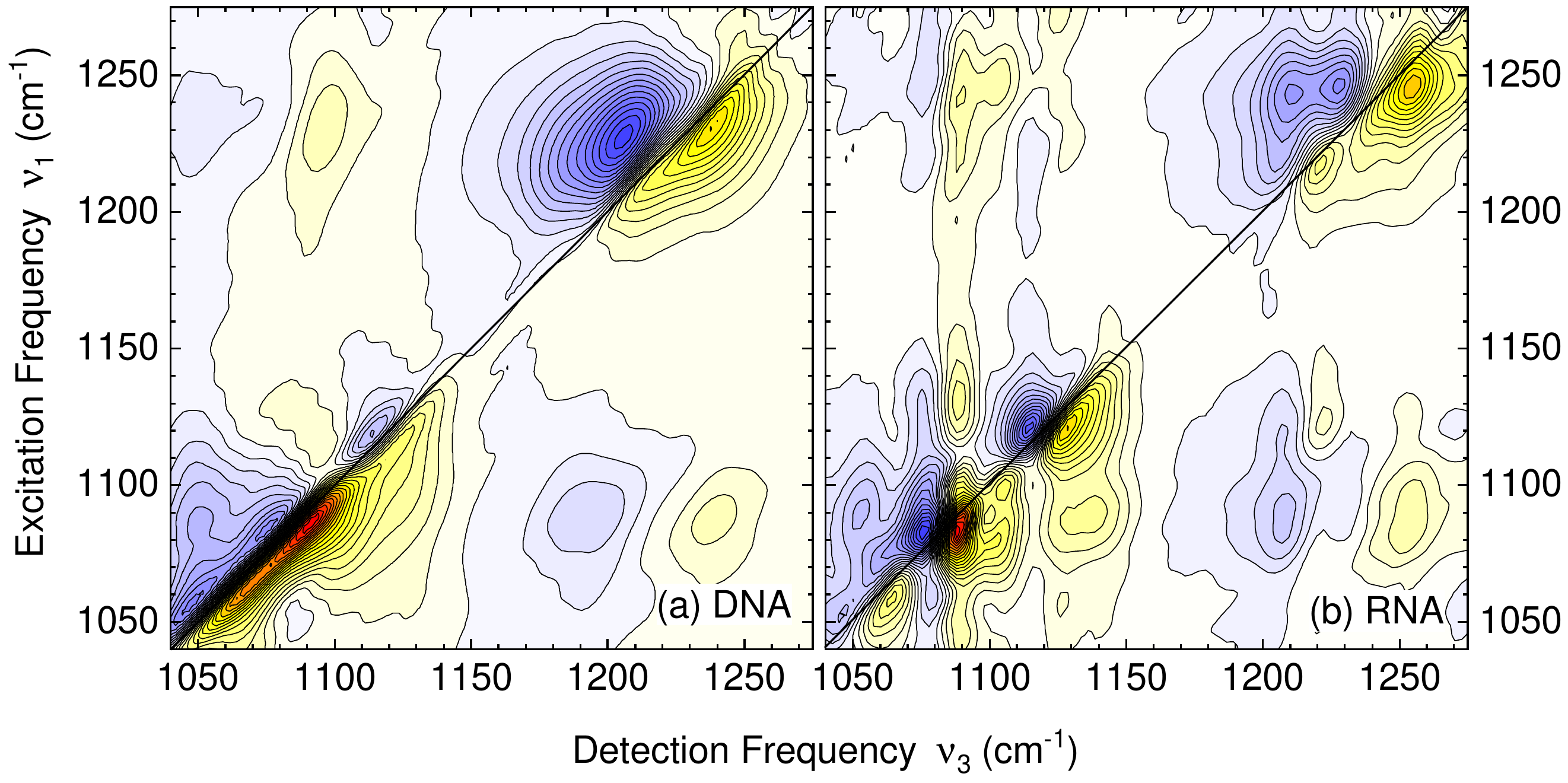}
%
%
\caption{Two-dimensional infrared (2D-IR) spectrum of (a) DNA oligomers containing 23 alternating adenine-thymine pairs at 92\% r.h., and (b) fully hydrated RNA oligomers containing 23 adenine-uracil pairs. The absorptive 2D signal is plotted as a function of excitation and detection frequency. The spectra were recorded at a waiting time of T=250 fs. The signal changes by 5\% between neighboring contour lines. }
\label{fig:2}       
\end{figure}

\section{Two-Dimensional Infrared Spectra of DNA and RNA at Different Hydration Levels}
\label{sec:3}

The 2D-IR spectra presented in the following were recorded in two partly overlapping spectral ranges, the frequency interval between 1050 and 1300 cm$^{-1}$ with the PO$_2^-$ stretching modes P1 and P2, and the range from 900 to 1150 cm$^{-1}$ in which the majority of the other backbone modes is located. The 2D-IR spectra of DNA and RNA oligomers shown in Fig\BF{.} 2 display  a series of peaks along the frequency diagonal $\nu_1 = \nu_3$, each consisting of a component due to absorption bleaching and stimulated emission on the fundamental v=0-1 transition (yellow-red contours) and of a contribution caused by the v=1-2 absorption (blue contours). The latter is anharmonically red-shifted to lower detection frequencies. The uppermost diagonal peaks in the range $(\nu_1,\nu_3) > (1200,1200)$ cm$^{-1}$ are due to the P1 modes. There is a single diagonal peak in the case of DNA (Fig. 2a) while the RNA spectrum in Fig. 2(b) displays two contributions, in agreement with the substructure observed in the linear infrared spectrum of Fig. 1(b). Below (1150,1150) cm$^{-1}$, both the DNA and RNA 2D spectra show a pronounced diagonal peak due to the P2 mode around (1090,1090) cm$^{-1}$ which in the case of RNA is complemented by peaks from the C$_{2'}$-OH stretching and the C$_{1'}$-O$_{4'}$-C$_{4'}$ ribose stretching vibrations (Fig. 2b). The diagonal 2D peaks in the DNA spectrum of Fig. 2(a) are of elliptic shapes, thus
revealing a substantial inhomogeneous broadening due to structural disorder of the hydrated helices. In contrast, inhomogeneous broadening is less present in the RNA 2D spectrum of Fig. 2(b), pointing to a more homogeneous hydration structure. The vibrational lifetimes of the different backbone modes are on the order of 1 ps, with the exception of the P1 mode. The latter shows a substantially shorter lifetime between 300 and 360 fs in all phosphate systems in aqueous environment studied by femtosecond vibrational spectroscopy so far. 

The pattern of cross peaks reflects the rich anharmonic coupling scheme between the backbone vibrations. First, the P1 and P2 modes of both DNA and RNA couple to each other, as is evident from the respective pairs of cross peaks. In RNA, there are no cross peaks between the two components of the P1 dublet, suggesting that the two contributions originate from different phosphate groups. The lower P1 component at 1220 cm$^{-1}$ couples strongly to the C$_{2'}$-OH stretching vibration, giving rise to the cross peaks at (1220,1120) and (1120,1220) cm$^{-1}$.

\begin{figure}[t]
\sidecaption[b]
\centering
\includegraphics[scale=1.1]{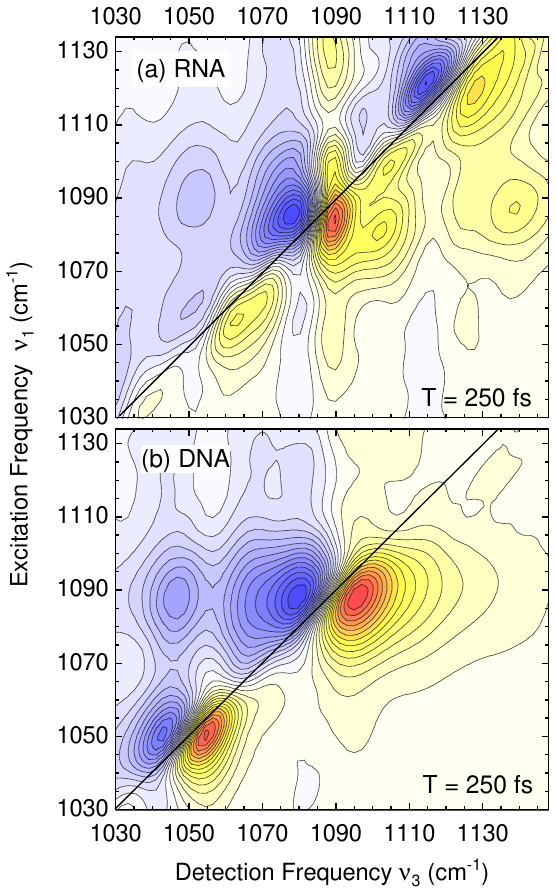}
%
%
\caption{Comparison of 2D-IR spectra of fully hydrated (a) RNA oligomers containing 23 adenine-uracil base pairs, and (b) DNA oligomers consisting of 23 alternating adenine-thymine base pairs. RNA displays a more complex mode and coupling pattern in the spectral range from 1030 to 1150 cm$^{-1}$. }
\label{fig:4}       
\end{figure}

Figure 3 shows a more detailed comparison of 2D-IR spectra of fully hydrated RNA and DNA oligomers between 1030 and 1150 cm$^{-1}$, corresponding to the lower frequency range of Fig. 2. The P2 diagonal peak of RNA (Fig. 3a) is oriented parallel to the $\nu_1$ axis, revealing a predominant homogeneous broadening which is in contrast to the inhomogeneously broadened P2 peak of DNA (Fig. 3b). The presence of the additional C$_{2'}$-OH stretching and the C$_{1'}$-O$_{4'}$-C$_{4'}$ ribose stretching vibrations gives rise to the broad and asymmmetric diagonal peak at (1130,1130)~cm$^{-1}$ and the complex cross peak pattern between the different modes. The reduced inhomogeneous broadening of RNA compared to DNA 2D peaks represents an important prerequisite for the observation of the fine details of vibrational coupling.

\begin{figure}[t]
\sidecaption
\centering
\includegraphics[scale=0.21]{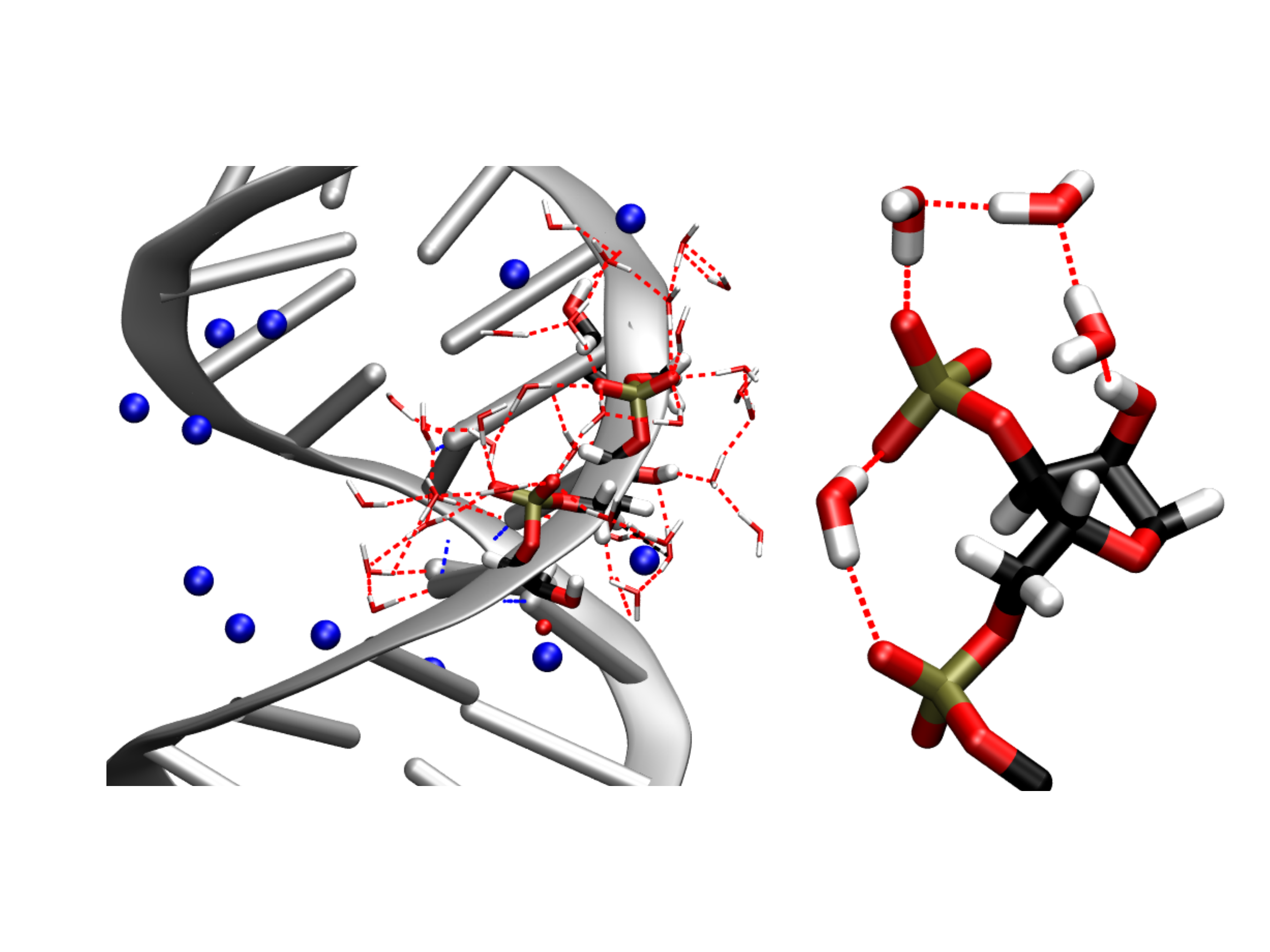}
%
%
\caption{Solvation structure of RNA oligomer backbone with counterions shown in blue. The magnification (right) highlights a PO$_2^-$...PO$_2^-$ bridging water molecule and a hydrogen bond network ring structure mediated by the C2'-OH group.}
\label{fig:bf1}       
\end{figure}

Figure~\ref{fig:bf1} depicts solvation structures  around double stranded RNA obtained from molecular dynamics simulations (cf. sec.~\ref{sec:MD}). 
The magnification (Fig.~\ref{fig:bf1}, right) highlights partially ordered water structures at the surface of RNA, i.e.,  water molecules bridging adjacent PO$_2^-$...PO$_2^-$ units and a hydrogen bond network ring structure mediated by the C\BF{$_{2'}$}-OH group. Such hydration structures are found in addition to  the solvation of phosphate groups by three water molecules  in an approximate tetrahedral geometry which is prototypical for DNA \cite{Schneider:BJ:1998,Kopka:JMB:1983}. Due to the A-helical geometry of RNA, the distances between neighboring phosphate groups are reduced compared to DNA in the B-form \cite{Saenger:Nature:1986}. Accordingly, the phosphate groups can be bridged by single water molecules that form strong hydrogen bonds with the PO$_2^-$ oxygen atoms. The additional OH group at the C2' position in RNA not only introduces additional vibrational modes but mediates hydrogen bonded rings of $\approx$ 3 water molecules to the PO$_2^-$ oxygen atoms. Such ring structures lead to partially ordered  water structure in the minor groove of RNA. Both, bridging water and ring structures observed in  the molecular dynamics simulations  are in line with reports from X-ray diffraction studies \cite{Egli:BP:1998,Egli:Biochem:1996} that suggest  that the first hydration layer around RNA contains a larger number of water molecules in a more ordered arrangement.

Compared to DNA, the partly ordered hydration structure around the backbone of  RNA imposes reduced heterogeneity of the first water layers around the helix. Such observations  corroborate the reduced inhomogeneous width observed for the 2D lineshapes of RNA (cf. Figs.~\ref{fig:2}, \ref{fig:4}). For example,  the P2 diagonal peak shows substantially reduced inhomogeneity and a pronounced coupling pattern to  C$_{2'}$-OH stretching and the C$_{1'}$-O$_{4'}$-C$_{4'}$ ribose stretching vibrations where the limited inhomogeneous broadening of the diagonal peaks reflects the  more ordered character of the hydration shell structure around RNA.The water molecules of the ring structures could be be instrumental in inducing the pronounced intermode vibrational couplings between the phosphate group and the ribose unit. Further, the doublet structure of P1 peaks and the absence of cross peaks for this feature suggests distinct different phosphate group hydration sites along the A-helix. The reduced inhomogeneous broadening here  is essential for separating the different P1 components and respective coupling patterns to the C$_{2'}$-OH stretching and C$_{1'}$-O$_{4'}$-C$_{4'}$ ribose stretching vibrations.

In the following, we focus on 2D-IR spectra of DNA oligomers with 23 
alternating A-T pairs at a hydration level of 92 \% r.h. This water content 
of the sample corresponds to 20 to 30 water molecules per base pair which is 
sufficient to form two closed water layers around the B-helix of the 
oligomer. Figures 5(a,b) show absorptive 2D spectra in the spectral range 
from 900 to 1150 cm$^{-1}$, recorded at waiting times of 500 and 2000 fs. The 
arrows and the symbols on the right hand side of Fig.~5(b) give assignments of 
the diagonal peaks to the different backbone modes (cf. Fig. 1a). \BF{I}n addition 
to the diagonal peaks, the spectra display a complex pattern of cross peaks, 
appearing as stripe-like features of alternating sign in the upper triangle 
of the 2D frequency plane. The cross peaks indicate anharmonic couplings between 
essentially all backbone modes.

\begin{figure}[t]
\centering
\includegraphics[scale=0.6]{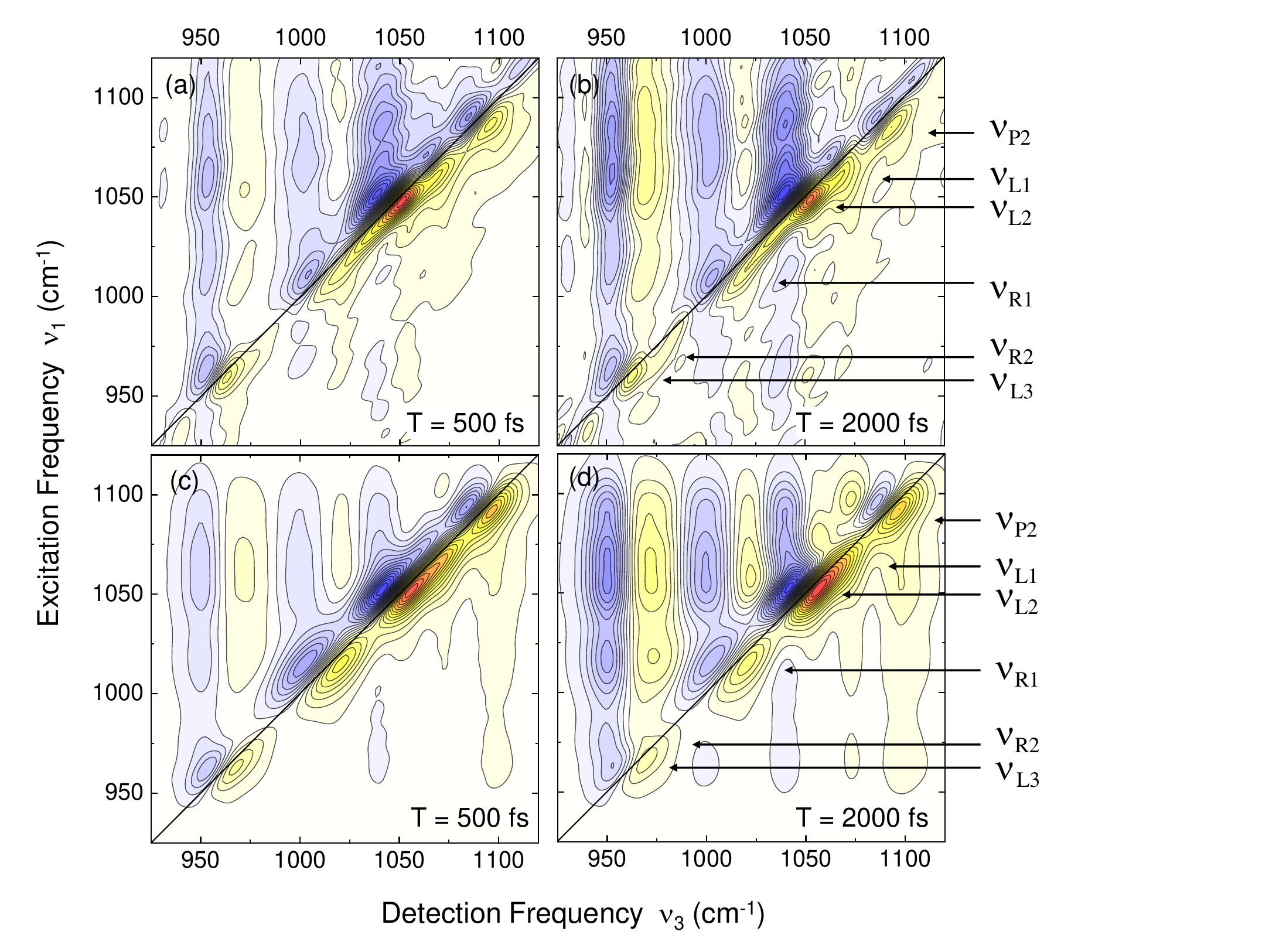}
%
%
\caption{Absorptive 2D-IR spectra of backbone modes of a double-stranded DNA oligomer containing 23 alternating adenine-thymine base pairs at a hydration level of 92\% r.h. (a,b) Experimental spectra recorded at waiting times T=500 fs and T=2000 fs. (c,d) Simulated spectra. The symbols and arrows on the right-hand side mark the different backbone modes contributing to the spectra. }
\label{fig:3}       
\end{figure}

The complex arrangement of diagonal and cross peaks is preserved with unchanged spectral positions at longer waiting times. The relative intensities of the different peaks, however, undergo substantial changes. In particular, the relative strengths of the cross peaks at low detection frequencies $\nu_3$ increases with waiting time. This behavior is a hallmark of inter-mode energy transfer, i.e., energy is transferred from the backbones modes at higher frequencies to those at low frequencies on a time scale of a few picoseconds. This process together with the vibrational relaxation of each individual mode determines the time evolution of the diagonal peaks as has been shown in more detail in ref.~\cite{Siebert:JPCB:2015}.

The experimental 2D-IR spectra were analyzed with the help of the simulation approach outlined in section 2.3.4. This treatment combines a density matrix approach for calculating the third-order nonlinear response to the photon-echo pulse sequence with a Kubo ansatz for the FFCF (cf. equation \ref{eq:ffcf}). For reproducing the experimental results, the decay times in the Kubo FFCF were set as $\tau_1=300$ fs and $\tau_2  =50$ ps, the latter leading
to a quasi-static inhomogeneous broadening of the peaks in the 2D-IR spectrum. With such fixed correlation times, the fluctuation amplitudes $\Delta_{1,2}$ were adjusted for each of the seven modes to reproduce the observed lineshapes. The diagonal anharmonicities of the modes which determine the separation of the two components of the diagonal peaks along the detection frequency $\nu_3$ were chosen for each mode individually while a global anharmonic intermode couplng of 10~cm$^{-1}$ was chosen in order to calculate the cross peak pattern. The model includes incoherent down-hill energy transfer between the modes with a time constant of 2 ps and population relaxation of the v=1 states of the vibrations with an average time constant of 1.7 ps. In parallel to the analysis of the 2D spectra, the linear infrared absorption spectrum of the DNA oligomers was fitted to benchmark the derived parameters entering in the Kubo FFCF.

%
%

%
%

\begin{figure}[t]
\sidecaption[b]
\centering
\includegraphics[scale=0.7]{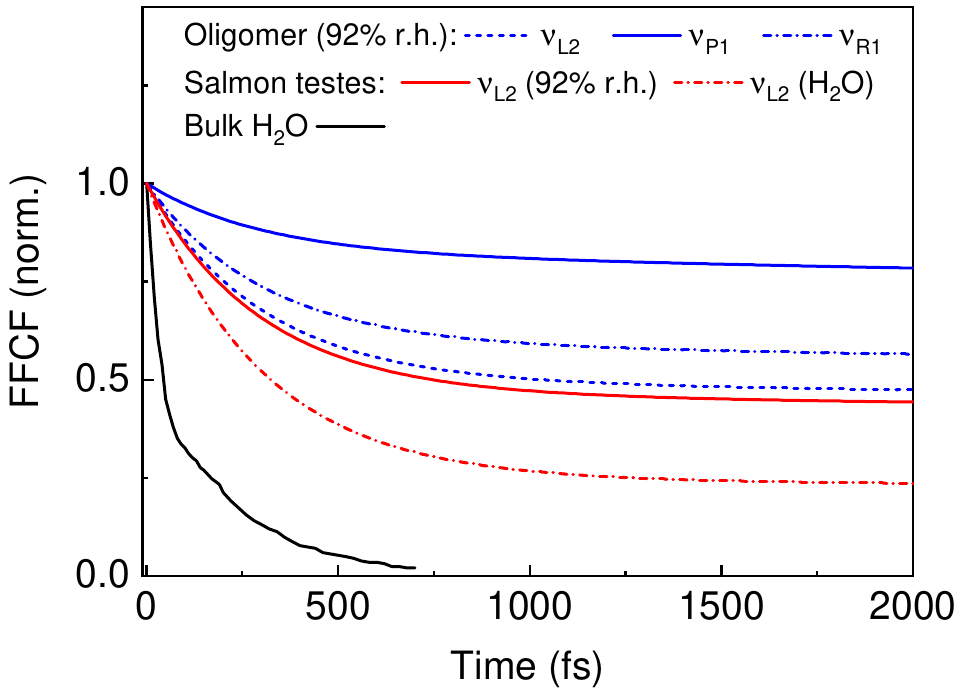}
%
%
\caption{Frequency fluctuation correlation functions (FFCFs) derived from the simulations of 2D-IR spectra of backbone modes of DNA oligomers and salmon testes DNA at different hydration levels. The FFCFs consist of a first term with a correlation decay time of 300 fs and a quasi-static contribution. The amplitudes of the two components are different for the different modes. Black solid line: FFCF of bulk water calculated in ref.~\cite{Jansen:JCP:2010} . }
\label{fig:5}       
\end{figure}

The calculated spectra presented in Figs. 5(c,d) are in quantitative agreement with the experimental results in Figs. 5(a,b). The diagonal anharmonicities are between 5 and 10 cm$^{-1}$, the fluctuation amplitudes in the Kubo function are $\Delta_1=6-8$ cm$^{-1}$ and $\Delta_2=6-16$ cm$^{-1}$. Selected FFCFs for the P1, L2, and R1 modes are plotted in Fig. 6 (blue lines), a detailed listing of the parameter values has been given in ref.~\cite{Siebert:JPCB:2015}. The initial 300 fs decay of the FFCF accounts for the antidiagonal width of the different 2D peaks which is determined by spectral diffusion of the vibrational transitions. Spectral diffusion mainly originates from the fluctuating Coulomb forces the first few water layers and, to lesser extent, solvated counterions exert on the backbone. Structural low-frequency fluctuations of the DNA backbone with its phosphate ions may contribute as well. The 300 fs decay is moderately slower than the initial sub-100 fs decay of the FFCF of bulk H$_2$O which is shown as a black solid line in Fig. 6 \cite{Jansen:JCP:2010}. Water motions at the interface to a DNA helix, in particular librational degrees of freedom, are restricted by the steric boundary conditions the DNA surface sets in its major and minor grooves. This results in a somewhat slower FFCF decay compared to bulk water. However, the first few layers of the water shell are far from representing a rigid structure and, thus, the concepts of an 'iceberg-like' water structure and/or extremely slow biological water can be ruled out safely \cite{Laage:ChemRev:2017}. 

The second component of the FFCFs with a decay time beyond 10 ps results in quasi-static inhomogeneous broadening as manifested in the elliptic shape of the 2D diagonal peaks. This behavior reflects a comparably long-lived structural heterogeneity of hydration geometries, i.e., structural disorder along the DNA backbone. Hydrogen bonds between the water molecules in the first hydration layer and phosphate groups in the backbone are stronger than water-water hydrogen bonds and display, according to detailed molecular simulations of the interface \cite{Duboue:JACS:2016}, lifetimes on the order of 20 ps. As a result, the molecular pattern of the first water layer is preserved for this picosecond time range and differences in the local number of waters along the helix are expected to result in a quasi-static distribution of vibrational frequencies of the phosphate groups and the other interacting units of the backbone. It should be recalled that inhomogeneous broadening is less pronounced in the 2D-IR spectra of RNA oligomers (cf. Figs. 2 and 3), resulting in systematically smaller fluctuation amplitudes $\Delta_2$ in the relevant FFCFs
\cite{Bruening:JPCL:2018}. Here, the additional OH group of the ribose unit has a strong impact on the first-layer water structure and induces an overall more homogeneous hydration pattern.

\begin{figure}[t]
\centering
\includegraphics[scale=0.63]{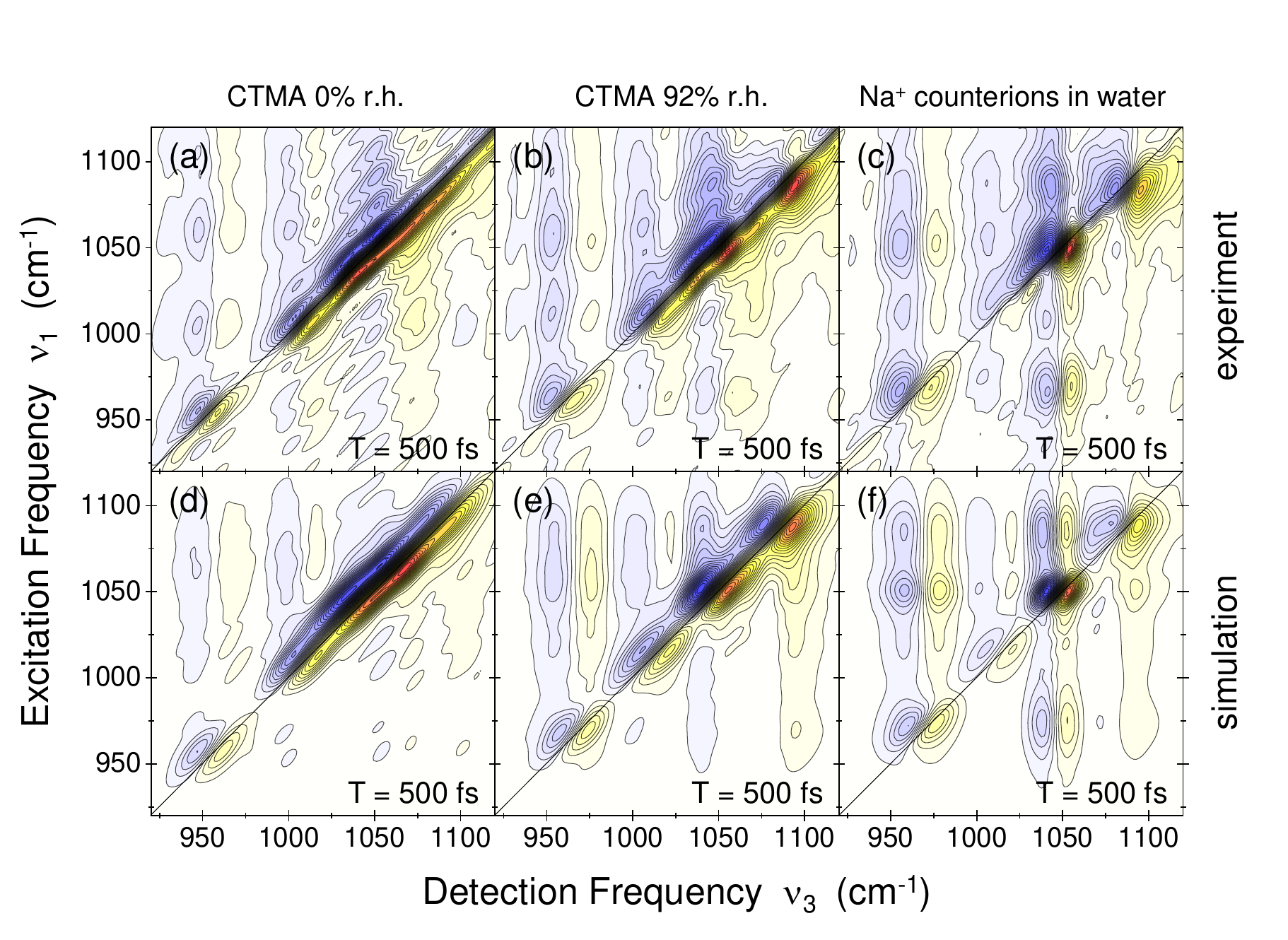}
%
%
\caption{2D-IR spectra of salmon testes DNA at different hydration levels. The spectra at 0\% and 92\% r.h. were measured with thin-film samples containing CTMA counterions. (a-c) Experimental spectra recorded at a waiting time T=500 fs. (d-f) Simulated spectra.
}
\label{fig:6}       
\end{figure}

\section{Electric Fields at the DNA Surface and Energy Exchange Processes}
\label{sec:4}

The combination of 2D-IR spectroscopy with ab-initio based theory allows for an in-depth analysis of the electric interactions between a DNA helix and its water shell. Key issues are the strength and fluctuation amplitude of the interfacial electric fields, the relative contributions of water molecules, counterions, and phosphate groups to the local fields, and the spatial range of the electric forces in the dense water shell. Such issues have been addressed in a series of theoretical studies, starting with early polyelectrolyte theories and developing towards descriptions at the atomic level based on Poisson Boltzmann pictures for static geometries and/or classical molecular dynamics simulations. In the latter, the choice of the particular water model represents a major issue and most recent work has implemented flexible polarizable water molecules, e.g., on the SPC level. While early simulation work has been limited to a time range up to a few nanosecond, current simulations have provided trajectories extending into the microsecond range.  

The backbone modes of DNA and RNA are most sensitive and noninvasive probes of interfacial electric fields. In the following, we present recent results from 2D-IR experiments with native salmon testes DNA at different hydration levels. Salmon testes DNA with its approximately 2000 base pairs displays the same set of backbone normal modes as the much shorter oligomers discussed so far and a similar infrared absorption spectrum (cf. Fig. 1a). The  experiments give insight into the fluctuation amplitudes of electric fields via the 2D line shapes and, in combination with theory, in the relevant strength and range of electric interactions.

Figure 7 summarizes absorptive 2D-IR spectra of salmon testes DNA at \BF{three} different hydration levels at a population time of T=500 fs. The spectra in Figs. 7(a,b) were recorded with thin-film samples containing CTMA instead of the genuine Na$^+$ counterions, while Fig. 7(c) shows a 2D-IR spectrum of fully hydrated DNA with Na$^+$ counterions. The basic pattern of diagonal and cross peaks is identical in all spectra and close to the 2D spectra of the much shorter DNA oligomers (cf. Fig. 5), with some minor spectral shifts and slight variations in the relative peak intensities. There are, however, systematic changes of the 2D line shapes with hydration which have been discussed in detail in ref.~\cite{Siebert:JPCL:2016}. The antidiagonal widths of the peaks on the frequency diagonal $\nu_1=\nu_3$ are practically identical for 92 \% r.h. (20-30 water molecules per base pair) and full hydration (more than 150 water molecules per base pair). In contrast, the 2D spectrum for 0 \% r.h. (2 waters per base pair) shows narrower profiles along the antidiagonals. An opposite trend is found for the spectral widths along the diagonal which decrease with increasing hydration. As a result, the envelopes of the diagonal peaks in the spectrum of the fully hydrated samples appear tilted towards the vertical frequency axis $\nu_1$, i.e., they appear more homogeneous. Additional measurements were performed with a fully hydrated sample in which the Na$^+$ counterions were exchanged against Mg$^{2+}$ ions. Within the experimental accuracy, the 2D-IR spectrum was identical to the one shown in Fig. 7(c).

The salmon testes 2D-IR spectra were analyzed with the formalism described in sections 2.3.4 and 3, again using correlation times of 300 fs and 50 ps in the Kubo ansatz for the FFCF \BF{(eq.~\ref{eq:ffcf})}. The calculated spectra shown in Figs. 7(d-f) are in good agreement with the experiments. The FFCFs of the L2 mode of salmon testes DNA at 92 \% r.h. and full hydration are included in Fig. 6 (red lines). The FFCF for 92\% r.h. agrees very well with the corresponding result for the DNA oligomers (blue dashed line) while the FFCF for full hydration exhibits a reduced amplitude of the quasi-static 50 ps component (red dash-dotted line), in line with the more homogeneous 2D lineshape of the L2 peak.
 
The fluctuation amplitudes for the 300 fs FFCF component for the different modes have values of $\Delta_1= 6 - 11$ cm$^{-1}$ which are the same for full hydration and 92 \% r.h. In contrast, values up to $\Delta_1=5$ cm$^{-1}$ are found at 0 \% r.h. We conclude that (i) the water shell represents the predominant source of the fluctuating electric force giving rise to spectral diffusion, and (ii) that this force originates essentially from the first two water layers which are present at both 92 \% r.h. and for full hydration. In other words, the dipolar water molecules are the main source of short-range Coulomb forces which cause the fast frequency modulation of the vibrational transitions. The fluctuation amplitudes $\Delta_1$ are proportional to the fluctuation amplitudes of the electric field from the two water layers, in the simplest linear approximation according to $\Delta_{1i} = a_i \cdot \Delta E_i$ where $\Delta E_i$ is the fluctuation amplitude projected on the vibrational coordinate of the respective backbone normal mode. The tuning rates $a_i$ of the vibrational transition frequencies can be estimated from molecular force fields for the different vibrations. Recent theoretical calculations for the symmetric (P2) and asymmetric (P1) PO$_2^-$ stretching vibrations suggest a value of $a_i =0.4 - 0.5$ cm$^{-1}$/(MV/cm). With such tuning rates, one derives a fluctuation amplitude $\Delta E_i = 20 - 25$ MV/cm with the help of the experimental $\Delta_{1,\rm{P}2}=10$ cm$^{-1}$.   

\begin{figure}[t]
\centering
\includegraphics[scale=0.25]{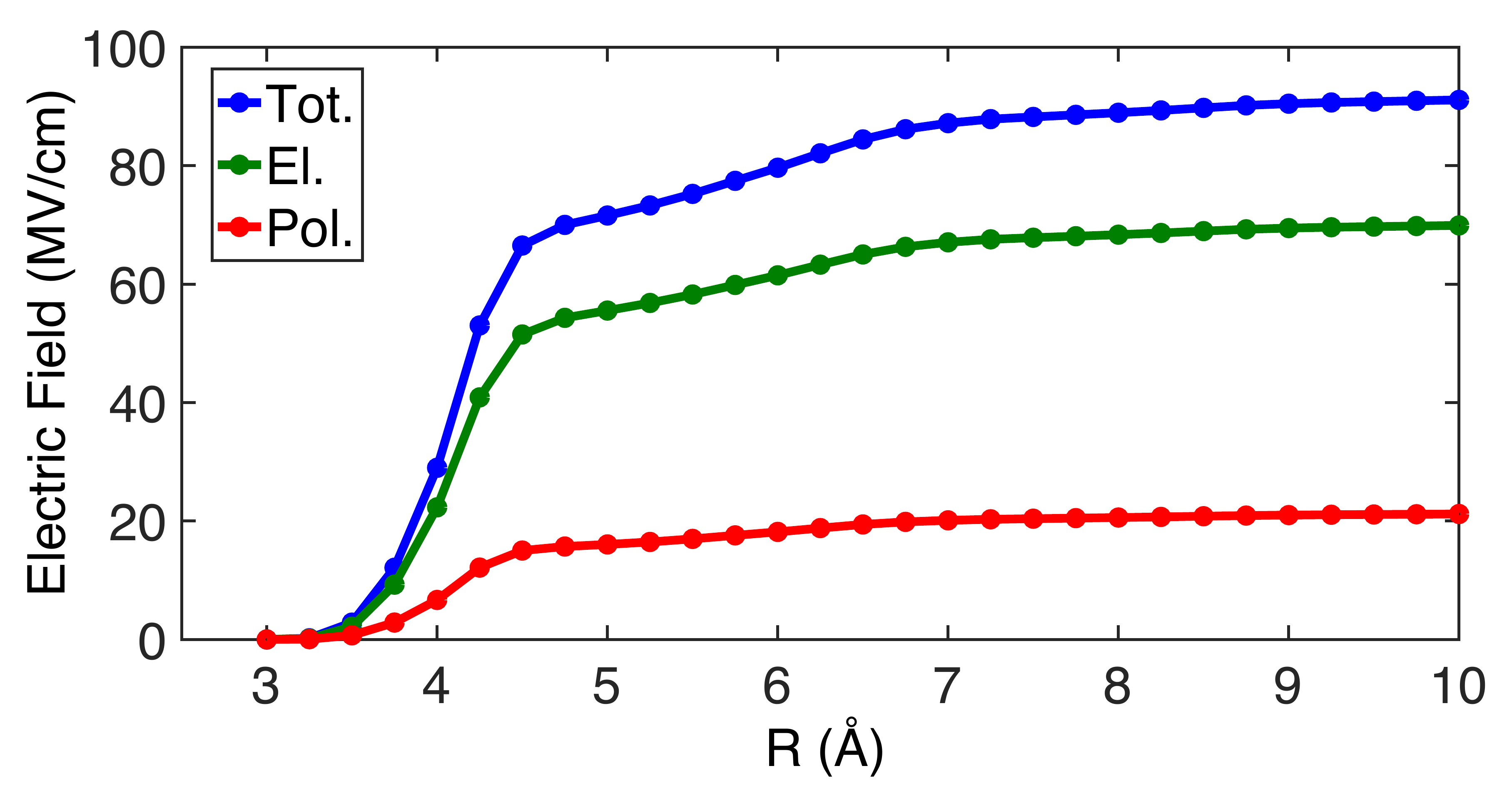}
%
%
\caption{Radial dependence of electric fields calculated for dimethyl phosphate (DMP) in bulk H$_2$O. The electric field strength projected on the C2 axis of the PO$_2^-$ group is plotted as a function of radial distance R in angstroms. The total field consists of a predominant electrostatic component and a contribution arising from the mutual polarization of DMP and the solvent.
}
\label{fig:7}       
\end{figure}

To benchmark the fluctuating electric forces relevant for spectral diffusion in the DNA 2D-IR spectra, theoretical calculations were performed for dimethyl phosphate [DMP, (CH$_3$)$_2$PO$_4^-$] in bulk H$_2$O. DMP has repeatedly been used as a model system for the phosphodiester moiety of the DNA and RNA backbones. The fluctuating electric forces on the phosphate group were calculated with the approach described in section 2.3\BF{.2} which includes both electric fields from static multipoles and polarizations due to induced dipoles.   

Figure 8 shows the dependence of the calculated time-averaged electric field from the water shell projected on the C2 axis of the PO$_2^-$ entity and integrated over an increasing radial distance R from the phosphate group of DMP. The field reaches a maximum value on the order of 100 MV/cm. The total electric field generated by water molecules arises to a large extent from the first water layer. The second layer contributes some 18 \% with noticeable contributions from induction. Furthermore, the solvent electric field experienced by the phosphate group is the dominant contribution to the pronounced solvatochromism of the asymmetric PO$_2^-$ stretching vibration. Accounting for a field expansion up to quadrupoles and polarization due to induced dipoles allows for simulating solvent induced frequency shifts and line shapes in almost quantitative agreement to experiment. The underlying field fluctuation amplitudes are found in the range $\Delta E \approx 25$ MV/cm, in excellent agreement with the experimental observations.  The model strongly supports the picture of short-range electric forces that arise locally from the first and second hydration shell.

The electric forces the hydration shell exerts on the DNA backbone depend sensibly on the spatial arrangement of water molecules in the first and second water layer. Changes of this arrangement have a direct impact on the frequency positions, absorption strengths, and lineshapes of the infrared absorption bands of backbone modes. In a recent pump-probe study \cite{Liu:StructDyn:2017}, the water shell was heated by a femtosecond infrared pulse which was resonant to its OH stretching absorption band. The decay of the OH stretching excitation and subsequent energy redistribution generate a heated ground state of the water shell in which the water-water hydrogen bond strengths and arrangement of water molecules is modified. This structure change modifies the infrared absorption of the backbone modes as was directly monitored by a femtosecond probe pulse. The rise of such absorption changes follows in time the buildup of the hot water ground state, i.e., the change of electric interactions due to the rearrangement of water molecules. Energy transfer from the heated water shell into the initially unexcited DNA helix occurs on a substantially slower time scale of tens of picoseconds.

The impact of electric forces  on transient line shapes of backbone modes is corroborated  by simulations of electric fields for the  DMP model system (cf. sec.~\ref{sec:field}) in bulk H$_2$O at elevated temperatures~\cite{Liu:StructDyn:2017}.
 While spatial rearrangements in the first and second solvation shell are moderate for increasing temperature, a resulting decrease of the electric field acting on the DMP phosphate group was found. The reduction of the field amplitudes induces a blue-shift of the symmetric and asymmetric PO$_2^-$ stretching vibrations  and is the origin of the observed modulation of line shapes. Such Coulomb-mediated coupling is predominantly mediated by the electric field the first few water layers exert on the backbone that changes upon formation of the hot water ground state. The limited reordering of water molecules in the first and second layer around the double helix demonstrates the pronounced sensitivity of the DNA backbone vibrations and their transient spectra to electric fields exerted by the  aqueous environment.

\section{Conclusions and Outlook}
\label{sec:5}

The results presented in this chapter demonstrate the important role of hydrogen bonds and Coulomb interactions for the structure and dynamics of biomolecules. Water molecules in the first hydration layer at the surface of DNA and RNA form hydrogen bonds with the phosphate groups in the backbone and with other polar groups of the backbone and the base pairs, resulting in a water structure different from the bulk liquid. The structure of the first hydration layer around RNA is affected by the additional OH group of the ribose units and less heterogeneous than around DNA. Local interactions at the hydrated interface and molecular dynamics connected with thermal excitations of low-frequency degrees of freedom are probed in a noninvasive way by backbone vibrations. The 2D-IR spectra of such modes display line shapes which are determined by spectral diffusion on a 300 fs time scale and by a quasi-static inhomogeneous broadening, reflecting the structural disorder along the helical DNA and RNA structures. Spectral diffusion is moderately slowed down compared to bulk water and originates mainly from fluctuating electric forces the water molecules in the first and second hydration layer exert on the different vibrational oscillators. Experiments at different hydration levels demonstrate a short-range character of electric interactions, typically extending over 2 water layers. The electric field strength at the hydrated interface reaches values on the order of 100 MV/cm with fluctuation amplitudes of some 25 MV/cm. An in-depth theoretical analysis for the hydrated model system dimethyl phosphate suggests a predominant electrostatic (dipole) contribution to the total electric field while polarization forces account for approximately 20 \% of the field at the phosphate groups.  

Counterions of the negatively charged DNA and RNA helices as well as positively charged excess ions are typically part of the native biomolecular environment. Their  interaction with the backbone, in particular with the ionic phosphate groups, plays an important role for the stabilization and folding of macromolecular structures. In this context, magnesium ions (Mg$^{2+}$) are particularly relevant. Ions located close to the DNA and RNA surfaces are expected to contribute to the interfacial electric fields while the role of ions solvated in the hydration shell is less obvious. Ions in the first solvation shell around phosphate groups  can form contact ion pairs whereas ions separated by at least one water layer from the backbone are partly mobile and part of the so-called diffuse ion atmosphere. 

Electric interactions with the different types of ions have been the subject of extensive theoretical work, addressing the radial ion distributions around helices and the resulting field strength at the surfaces of DNA and RNA. In contrast, experimental work has remained limited. X-ray diffraction from crystallized samples has identified sites of essentially immobile ions in the vicinity of the DNA and RNA backbone but gives not much insight in the diffuse
ion atmosphere \cite{Vlieghe:AC:1999,Egli:BP:1998}. Small angle x-ray scattering has provided information on the thickness of the counterion layer and, to some extent, the radial concentration gradient \cite{Das:PRL:2003}. 
\begin{figure}[t]
\centering
\includegraphics[scale=0.5]{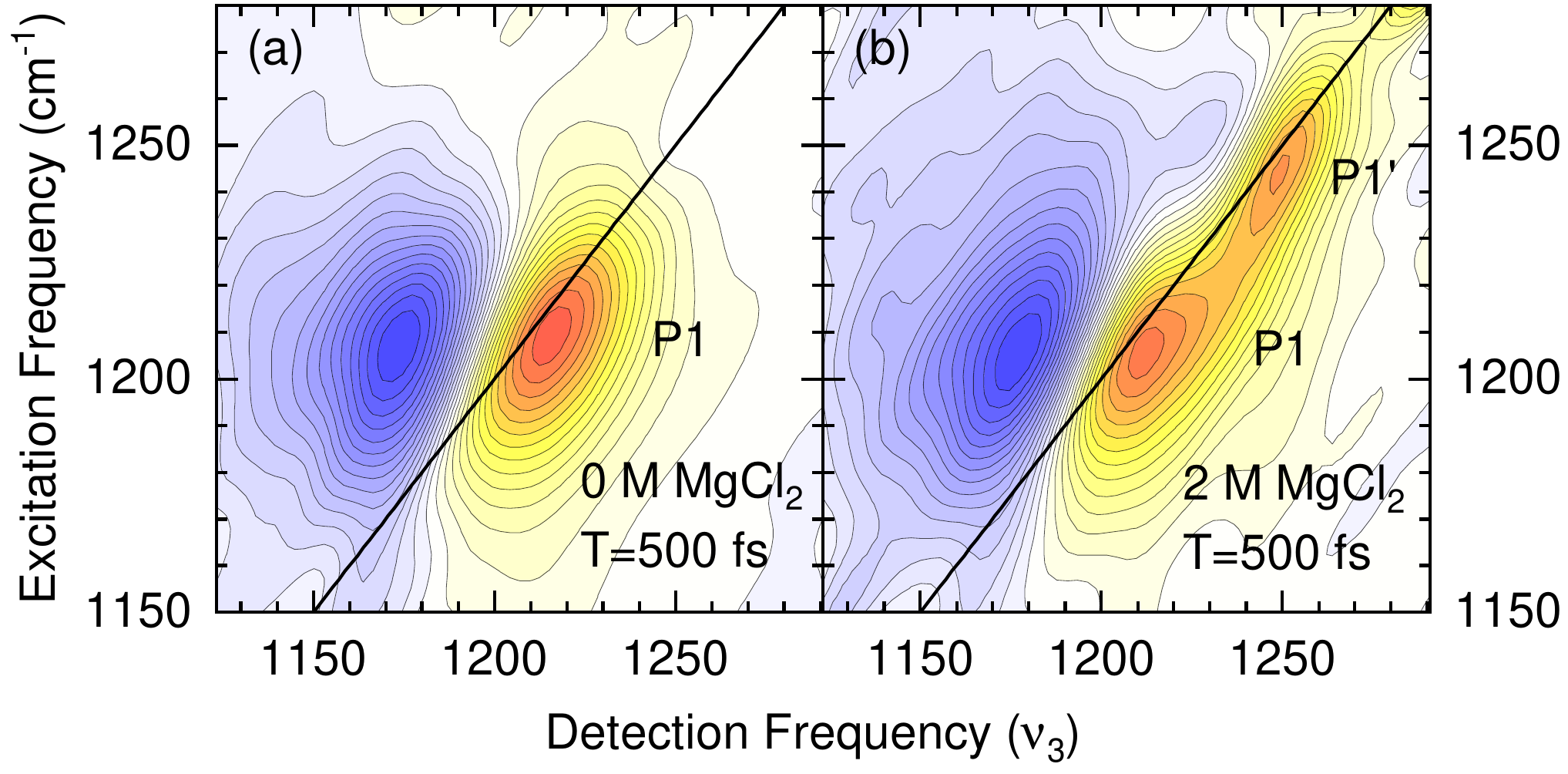}
%
%
\caption{Absorptive 2D-IR spectra of (a) dimethyl phosphate with sodium counterions (DMP-Na) in H$_2$O and (b) DMP-Na and
2 M Mg(H$_2$O)$_6$Cl$_2$ in H$_2$O. The spectra were recorded at a waiting time T=500 fs. The 2D signal difference between
neighboring contour lines is 7.5\%. }
\label{fig:10}       
\end{figure}

Very recently, we have studied the interaction of Mg$^{2+}$ ions with phosphate groups by 2D-IR spectroscopy of the model system sodium dimethyl phosphate (DMP$^-$Na$^+$) in an H$_2$O environment where DMP$^-$ and Na$^+$ are solvated separately \cite{Schauss:JPCL:2019}. The asymmetric PO$_2^-$ stretching vibration of DMP$^-$ serves as a probe. The 2D-IR spectrum in Fig. 9(a) measured with with a DMP$^-$ concentration of 0.2 M in absence of Mg$^{2+}$ ions displays the diagonal peak of the P1 vibration around (1210,1210) cm$^{-1}$, consisting of a component due to ground state bleaching and stimulated emission on the fundamental v=0-1 transition (yellow-red contours) and the red-shifted peak of v=1-2 absorption (blue contours). Upon addition of 2 M   Mg(H$_2$O)$_6$Cl$_2$, i.e., a tenfold excess of Mg$^{2+}$ ions, one observes a separate blue-shifted diagonal peak P1' around (1250,1250) cm$^{-1}$. It is important to note that there are no cross peaks between P1 and P1', pointing to their origin from different DMP$^{-}$ entities. Spectra recorded for different waiting times T and complementary pump-probe measurements give a decay time of 300 fs for P1 and of 400 fs for P1'. The two different species are also present in the linear infrared absorption spectra which were analyzed for different relative concentrations of DMP$^{-}$ and Mg$^{2+}$. 

An analysis of the relative concentrations of DMP$^-$, Mg$^{2+}$, their counterions Na$^+$ and Cl$^-$, and water shows that there is a significant fraction of DMP$^-$/Mg$^{2+}$ contact ion pairs, i.e., DMP$^-$ populates a site in the comparably rigid first hydration layer around Mg$^{2+}$ which consists (in absence of DMP$^-$) of 6 water molecules in an octahedral geometry. The strong interaction between the two ions in contact induces a blue-shift of the asymmetric PO$_2^-$ stretching vibration of the phosphate group. The microscopic origin of this behavior was identified in \BF{ab-initio simulations (sec.~\ref{sec:abinitio}) on} DMP$^-$(H\BF{$_2$}O)$_x$Mg$^{2+}$ clusters with x=19 water molecules. For large distances, i.e., three water layers between Mg$^{2+}$ and DMP$^-$, the frequency of the P1 mode is close to its value without any magnesium ions. Upon reducing the distance between the two ions, the P1 mode first displays a red-shift which is due to a spatial orientation of water molecules between the ions. The oriented water molecules exert a stronger electric field on the P1 oscillator, thus inducing the red-shift of its fundamental (v=0-1) frequency. For even shorter distances, with the DMP$^-$ and Mg$^{2+}$ ions in contact, the repulsive part of the ion-ion interaction prevails and induces a pronounced blue-shift of the P1 fundamental frequency. The calculated frequency shifts are in good agreement with the experimental results.

The application of phosphate vibrations as probes for contact ion pairs and, potentially, other ion configurations holds strong potential for analyzing the ion distribution around DNA and RNA on a multitude of time scales, ranging from hundreds of femtoseconds up to the micro- and millisecond time range of folding processes and other slower structure changes. Such work is presently underway.

\bibliography{BibtexReferenzen}

\begin{thebibliography}{10}
\providecommand{\url}[1]{{#1}}
\providecommand{\urlprefix}{URL }
\expandafter\ifx\csname urlstyle\endcsname\relax
  \providecommand{\doi}[1]{DOI \discretionary{}{}{}#1}\else
  \providecommand{\doi}{DOI \discretionary{}{}{}\begingroup
  \urlstyle{rm}\Url}\fi

\bibitem{Bakker:ChemRev:2010}
H.J. Bakker, J.L. Skinner, Chem. Rev. \textbf{110}, 1498 (2010)

\bibitem{Nibbering:ChemRev:2004}
E.T.J. Nibbering, T.~Elsaesser, Chem. Rev. \textbf{104}, 1887 (2004)

\bibitem{Laage:ChemRev:2017}
D.~Laage, T.~Elsaesser, J.T. Hynes, Chem. Rev. \textbf{117}, 10694 (2017)

\bibitem{Silvestrelli:PRL:1999}
P.L. Silvestrelli, M.~Parrinello, Phys. Rev.Lett. \textbf{82}, 3308 (1999)

\bibitem{Laage:Science:2006}
D.~Laage, J.T. Hynes, Science \textbf{311}, 832 (2006)

\bibitem{Ashihara:JPCA:2007}
S.~Ashihara, N.~Huse, A.~Espagne, E.T.J. Nibbering, T.~Elsaesser, J. Phys.
  Chem. A \textbf{111}, 743 (2007)

\bibitem{Saenger:Nature:1986}
W.~Saenger, W.N. Hunter, O.~Kennard, Nature \textbf{324}, 385 (1986)

\bibitem{Drew:JMB:1981}
H.R. Drew, R.E. Dickerson, J. Mol. Biol. \textbf{151}, 535 (1981)

\bibitem{Vlieghe:AC:1999}
D.~Vlieghe, J.P. Turkenburg, L.~van Meervelt, Acta Cryst. D: Biol. Cryst.
  \textbf{55}, 1495 (1999)

\bibitem{Schneider:BJ:1998}
B.~Schneider, K.~Patel, H.M. Berman, Biophys. J. \textbf{75}, 2422 (1998)

\bibitem{Egli:BP:1998}
M.~Egli, V.~Tereshko, M.~Teplova, G.~Minasov, A.~Joachimiak, R.~Sanishvili,
  C.M. Weeks, R.~Miller, M.A. Maier, H.~An, P.D. Cook, M.~Manoharan,
  Biopolymers \textbf{48}, 234 (1998)

\bibitem{Pal:ChemRev:2004}
S.K. Pal, A.H. Zewail, Chem. Rev. \textbf{104}, 2099 (2004)

\bibitem{Zhong:CPL:2011}
D.~Zhong, S.K. Pal, A.H. Zewail, Chem. Phys. Lett. \textbf{503}, 1 (2011)

\bibitem{Andreatta:2005aa}
D.~Andreatta, J.L. P{\'e}rez~Lustres, S.A. Kovalenko, N.P. Ernsting, C.J.
  Murphy, R.S. Coleman, M.A. Berg, J. Am. Chem. Soc. \textbf{127}, 7270 (2005)

\bibitem{Halle:JPCB:2009}
B.~Halle, L.~Nilsson, J. Phys. Chem. B \textbf{113}, 8210 (2009)

\bibitem{Furse:2008aa}
K.E. Furse, S.A. Corcelli, J. Am. Chem. Soc. \textbf{130}, 13103 (2008)

\bibitem{Yang:JPCB:2011}
M.~Yang, {\L}.~Szyc, T.~Elsaesser, J. Phys. Chem. B \textbf{115}, 13093 (2011)

\bibitem{Siebert:JPCB:2015}
T.~Siebert, B.~Guchhait, Y.~Liu, R.~Costard, T.~Elsaesser, J. Phys. Chem. B
  \textbf{119}, 9670 (2015)

\bibitem{HammZanni}
P.~Hamm, M.~Zanni, \emph{{Concepts and methods of 2D infrared spectroscopy}}
  (Cambridge University Press, 2011)

\bibitem{ChoBook}
M.~Cho, \emph{Two-dimensional optical spectroscopy} (CRC Press, Boca Raton
  2009, 2009)

\bibitem{Guchhait:StructDyn:2016}
B.~Guchhait, Y.~Liu, T.~Siebert, T.~Elsaesser, Struct. Dyn. \textbf{3}, 043202
  (2016)

\bibitem{Siebert:JPCL:2016}
T.~Siebert, B.~Guchhait, Y.~Liu, B.P. Fingerhut, T.~Elsaesser, J. Phys. Chem.
  Lett. \textbf{7}, 3131 (2016)

\bibitem{Liu:StructDyn:2017}
Y.~Liu, B.~Guchhait, T.~Siebert, B.F. Fingerhut, T.~Elsaesser, Struct. Dyn.
  \textbf{4}, 044015 (2017)

\bibitem{Costard:JCP:2015}
R.~Costard, T.~Tyborski, B.P. Fingerhut, T.~Elsaesser, J. Chem. Phys.
  \textbf{142}, 212406 (2015)

\bibitem{Costard:PCCP:2015}
R.~Costard, T.~Tyborski, B.P. Fingerhut, Phys. Chem. Chem. Phys. \textbf{17},
  29906 (2015)

\bibitem{Fingerhut:JCP:2016}
B.P. Fingerhut, R.~Costard, T.~Elsaesser, J. Chem. Phys. \textbf{145}, 115101
  (2016)

\bibitem{Tanaka:JACS:1996}
K.~Tanaka, Y.~Okahata, J. Am. Chem. Soc. \textbf{118}, 10679 (1996)

\bibitem{Dwyer:RSI:2013}
J.R. Dwyer, {\L}.~Szyc, E.T.J. Nibbering, T.~Elsaesser, Rev. Sci. Instrum.
  \textbf{84}, 036101 (2013)

\bibitem{Falk:JACS:1963}
M.~Falk, K.A. Hartman, R.C. Lord, J. Am. Chem. Soc. \textbf{85}, 387 (1963)

\bibitem{Banyay:BiophysChem:2003}
M.~Banyay, M.~Sarkar, A.~Gr{\"a}slund, Biophys. Chem. \textbf{104}, 477 (2003)

\bibitem{Guan:Biopol:1996}
Y.~Guan, G.J. Thomas, Biopolymers \textbf{39}, 813 (1996)

\bibitem{Bruening:JPCL:2018}
E.M. Bruening, J.~Schauss, T.~Siebert, B.P. Fingerhut, T.~Elsaesser, J. Phys.
  Chem. Lett. \textbf{9}, 583 (2018)

\bibitem{Cowan:Nature:2005}
M.L. Cowan, B.D. Bruner, N.~Huse, J.R. Dwyer, B.~Chugh, E.T.J. Nibbering,
  T.~Elsaesser, R.J.D. Miller, Nature \textbf{434}, 199 (2005)

\bibitem{Khalil:JPCA:2003}
M.~Khalil, N.~Demird{\"o}ven, A.~Tokmakoff, J. Phys. Chem. A \textbf{107}, 5258
  (2003)

\bibitem{Kirmizialtin_BiophysJ_2012}
S.~Kirmizialtin, A.R.J. Silalahi, R.~Elber, M.O. Fenley, Biophys. J.
  \textbf{102}, 829 (2012)

\bibitem{Young_JPCB_1998}
M.A. Young, B.~Jayaram, D.L. Beveridge, J. Phys. Chem. B \textbf{102}, 7666
  (1998)

\bibitem{Jayram_Biopol_1989}
B.~Jayaram, K.A. Sharp, B.~Honig, Biopolymers \textbf{28}, 975 (1989)

\bibitem{Cuervo_PNAS_2014}
A.~Cuervo, P.D. Dans, J.L. Carrascosa, M.~Orozco, G.~Gomila, L.~Fumagalli,
  Proc. Nat. Acad. Sci. U. S. A. \textbf{111}, E3624 (2014)

\bibitem{Ren:QRevBiophys:2012}
P.~Ren, J.~Chun, D.G. Thomas, M.J. Schnieders, M.~Marucho, J.~Zhang, N.A.
  Baker, Quart. Rev. Biophys. \textbf{45}, 427 (2012)

\bibitem{Cisneros_ChemRev_2014}
G.A. Cisneros, M.~Karttunen, P.~Ren, C.~Sagui, Chem. Rev. \textbf{114}, 779
  (2014)

\bibitem{Dans:CurOp:2016}
P.D. Dans, J.~Walther, H.~G{\'o}mez, M.~Orozco, Curr. Opinion Struct. Biol.
  \textbf{37}, 29  (2016)

\bibitem{Bevergidge:BiophysJ:2004}
D.L. Beveridge, G.~Barreiro, K.S. Byun, D.A. Case, T.E. Cheatham, S.B. Dixit,
  E.~Giudice, F.~Lankas, R.~Lavery, J.H. Maddocks, R.~Osman, E.~Seibert,
  H.~Sklenar, G.~Stoll, K.M. Thayer, P.~Varnai, M.A. Young, Biophys. J.
  \textbf{87}, 3799  (2004)

\bibitem{Perez:BiophysJ:2007}
A.~P{\'e}rez, I.~March{\'a}n, D.~Svozil, J.~Sponer, T.E. Cheatham, C.A.
  Laughton, M.~Orozco, Biophys. J. \textbf{92}, 3817  (2007)

\bibitem{Ivani:NatMeth:2015}
I.~Ivani, P.D. Dans, A.~Noy, A.~P{\'e}rez, I.~Faustino, A.~Hospital,
  J.~Walther, P.~Andrio, R.~Go{\~n}i, A.~Balaceanu, G.~Portella, F.~Battistini,
  J.L. Gelp{\'\i}, C.~Gonz{\'a}lez, M.~Vendruscolo, C.A. Laughton, S.A. Harris,
  D.A. Case, M.~Orozco, Nature Methods \textbf{13}, 55 EP  (2015)

\bibitem{Hart:JCTC:2012}
K.~Hart, N.~Foloppe, C.M. Baker, E.J. Denning, L.~Nilsson, A.D. MacKerell, J.
  Chem. Theor. Comput. \textbf{8}, 348 (2012)

\bibitem{Galindo-Murillo:NatCom:2014}
R.~Galindo-Murillo, D.R. Roe, T.E. Cheatham~III, Nature Commun. \textbf{5},
  5152 EP  (2014)

\bibitem{Rodrigo:JCTC:2016}
R.~Galindo-Murillo, J.C. Robertson, M.~Zgarbov{\'a}, J.~{\v S}poner,
  M.~Otyepka, P.~Jure{\v c}ka, T.E. Cheatham, J. Chem. Theor. Comp.
  \textbf{12}, 4114 (2016)

\bibitem{Adam:NucAcidRes:2017}
P.D. Dans, I.~Ivani, A.~Hospital, G.~Portella, C.~Gonz{\'a}lez, M.~Orozco,
  Nucl. Acids Res. \textbf{45}, 4217 (2017)

\bibitem{Sponer:JPCL:2014}
J.~{\v S}poner, P.~Ban{\'a}{\v s}, P.~Jure{\v c}ka, M.~Zgarbov{\'a},
  P.~K{\"u}hrov{\'a}, M.~Havrila, M.~Krepl, P.~Stadlbauer, M.~Otyepka, J. Phys.
  Chem. Lett. \textbf{5}, 1771 (2014)

\bibitem{Sponer:ChemRev:2018}
J.~{\v S}poner, G.~Bussi, M.~Krepl, P.~Ban{\'a}{\v s}, S.~Bottaro, R.A. Cunha,
  A.~Gil-Ley, G.~Pinamonti, S.~Poblete, P.~Jure{\v c}ka, N.G. Walter,
  M.~Otyepka, Chem. Rev. \textbf{118}, 4177 (2018)

\bibitem{Allner:JCTC:2012}
O.~Alln{\'e}r, L.~Nilsson, A.~Villa, J. Chem. Theor. Comput. \textbf{8}, 1493
  (2012)

\bibitem{Yoo:JPCL:2012}
J.~Yoo, A.~Aksimentiev, J. Phys. Chem. Lett. \textbf{3}, 45 (2012)

\bibitem{Martinek:JCP:2018}
T.~Martinek, E.~Dubou{\'e}-Dijon, {\v S}.~Timr, P.E. Mason, K.~Baxov{\'a}, H.E.
  Fischer, B.~Schmidt, E.~Pluha{\v r}ov{\'a}, P.~Jungwirth, J. Chem. Phys.
  \textbf{148}, 222813 (2018)

\bibitem{Li:ChemRev:2017}
P.~Li, K.M. Merz, Chem. Rev. \textbf{117}, 1564 (2017)

\bibitem{Lavery:NucAcidRes:2014}
R.~Lavery, J.H. Maddocks, M.~Pasi, K.~Zakrzewska, Nucl. Acids Res. \textbf{42},
  8138 (2014)

\bibitem{Salomon-Ferrer:JCTC:2013}
R.~Salomon-Ferrer, A.W. G{\"o}tz, D.~Poole, S.~Le~Grand, R.C. Walker, J. Chem.
  Theor. Comput. \textbf{9}, 3878 (2013)

\bibitem{Pall:CPC:2013}
S.~P{\'a}ll, B.~Hess, Comp. Phys. Commun. \textbf{184}, 2641 (2013)

\bibitem{Ren_JCPB_2003}
P.~Ren, J.W. Ponder, J. Phys. Chem. B \textbf{107}, 5933 (2003)

\bibitem{Troester_JPCB_2013}
P.~Tr{\"o}ster, K.~Lorenzen, P.~Tavan, J. Phys. Chem. B \textbf{118}, 1589
  (2014)

\bibitem{Ponder_JPCB_2010}
J.W. Ponder, C.~Wu, P.~Ren, V.S. Pande, J.D. Chodera, M.J. Schnieders,
  I.~Haque, D.L. Mobley, D.S. Lambrecht, J.~Robert A.~DiStasio, M.~Head-Gordon,
  G.N.I. Clark, M.E. Johnson, T.~Head-Gordon, J. Phys. Chem. B \textbf{114},
  2549 (2010)

\bibitem{Kuo_JPCB_2001}
I.~Kuo, , D.J. Tobias, J. Phys. Chem. B \textbf{105}, 5827 (2001)

\bibitem{Ploetz_JCTC_2016}
E.A. Ploetz, A.S. Rustenburg, D.P. Geerke, P.E. Smith, J. Chem. Theor. Comput.
  \textbf{12}, 2373 (2016)

\bibitem{Cisneros_ChemRev_2016}
G.A. Cisneros, K.T. Wikfeldt, L.~Ojam{\"a}e, J.~Lu, Y.~Xu, H.~Torabifard, A.P.
  Bart{\'o}k, G.~Cs{\'a}nyi, V.~Molinero, F.~Paesani, Chem. Rev. \textbf{116},
  7501 (2016)

\bibitem{Day_JCP_1996}
P.N. Day, J.H. Jensen, M.S. Gordon, S.P. Webb, W.J. Stevens, M.~Krauss,
  D.~Garmer, H.~Basch, D.~Cohen, J. Chem. Phys. \textbf{105}, 1968 (1996)

\bibitem{Gordon_JCPA_2001}
M.S. Gordon, M.A. Freitag, P.~Bandyopadhyay, J.H. Jensen, V.~Kairys, W.J.
  Stevens, J. Phys. Chem. A \textbf{105}, 293 (2001)

\bibitem{Gordon_ChemRev_2012}
M.S. Gordon, D.G. Fedorov, S.R. Pruitt, L.V. Slipchenko, Chem. Rev.
  \textbf{112}, 632 (2012)

\bibitem{Kaliman_JCC_2013}
I.A. Kaliman, L.V. Slipchenko, J. Comput. Chem. \textbf{34}, 2284 (2013)

\bibitem{Levinson2011}
N.M. Levinson, E.E. Bolte, C.S. Miller, S.A. Corcelli, S.G. Boxer, J. Am. Chem.
  Soc. \textbf{133}, 13236 (2011)

\bibitem{Fried_AccChemRes_2015}
S.D. Fried, S.G. Boxer, Acc. Chem. Res. \textbf{48}, 998 (2015)

\bibitem{Kim:ChemRev:2013}
H.~Kim, M.~Cho, Chem. Rev. \textbf{113}, 5817 (2013)

\bibitem{Basiak:AccChemRes:2017}
B.~B{\l}asiak, C.H. Londergan, L.J. Webb, M.~Cho, Acc. Chem. Res. \textbf{50},
  968 (2017)

\bibitem{Schauss:JPCL:2019}
J.~Schauss, F.~Dahms, B.F. Fingerhut, T.~Elsaesser, J. Phys. Chem. Lett.
  \textbf{10}, 238 (2019)

\bibitem{MukamelBible}
S.~Mukamel, \emph{{Principles of Nonlinear Optical Spectroscopy (Oxford Series
  on Optical and Imaging Sciences)}}, 3rd edn. (Oxford University Press, USA,
  1999)

\bibitem{Kopka:JMB:1983}
M.L. Kopka, A.V. Fratini, H.R. Drew, R.E. Dickerson, J. Mol. Biol.
  \textbf{163}, 129 (1983)

\bibitem{Egli:Biochem:1996}
M.~Egli, S.~Portmann, N.~Usman, Biochemistry \textbf{35}, 8489 (1996)

\bibitem{Jansen:JCP:2010}
T.l.C. Jansen, B.~Auer, M.~Yang, J.L. Skinner, J. Chem. Phys. \textbf{132},
  224503 (2010)

\bibitem{Duboue:JACS:2016}
E.~Duboue-Dijon, A.C. Fogarty, J.T. Hynes, D.~Laage, J. Am. Chem. Soc.
  \textbf{138}, 7610 (2016)

\bibitem{Das:PRL:2003}
R.~Das, T.T. Mills, L.W. Kwok, G.S. Maskel, I.S. Millett, S.~Doniach, K.D.
  Finkelstein, D.~Herschlag, L.~Pollack, Phys. Rev. Lett. \textbf{90}, 188103
  (2003)

\end{thebibliography}

\end{document}